\documentclass[iop,twocolappendix]{emulateapj}
\usepackage{natbib}
\usepackage{color}
\usepackage{mathptmx}
\usepackage{amsmath}
\usepackage{url}
\usepackage{multirow}
\usepackage{verbatim}  
\usepackage[breaklinks=true]{hyperref} 
\def\lsim{\lower.5ex\hbox{$\; \buildrel < \over \sim \;$}}
\def\gsim{\lower.5ex\hbox{$\; \buildrel > \over \sim \;$}}
\def\msun{\,{\rm M_\odot}}

\def\apj{ApJ}
\def\apjs{ApJS}
\def\araa{ARAA}
\def\mnras{MNRAS}
\def\nat{Nature}

\def\aj{AJ}
\def\apjl{ApJL}

\def\aap{Astronomy \& Astrophysics}

\begin{document}

\shorttitle{{\it AGORA} Galaxy Simulations Comparison}
\shortauthors{J. Kim et al. for the {\it AGORA} Collaboration}

\title{The {\it AGORA} High-Resolution Galaxy Simulations Comparison Project}

\author{Ji-hoon Kim\altaffilmark{1}}
\author{Tom Abel\altaffilmark{2}}
\author{Oscar Agertz\altaffilmark{3,4}}
\author{Greg L. Bryan\altaffilmark{5}} 
\author{Daniel Ceverino\altaffilmark{6}}
\author{Charlotte Christensen\altaffilmark{7}}
\author{Charlie Conroy\altaffilmark{1}}
\author{Avishai Dekel\altaffilmark{8}}
\author{Nickolay Y. Gnedin\altaffilmark{3,9,10}}
\author{Nathan J. Goldbaum\altaffilmark{1}} 
\author{Javiera Guedes\altaffilmark{11}}
\author{Oliver Hahn\altaffilmark{11}}
\author{Alexander Hobbs\altaffilmark{11}}
\author{Philip F. Hopkins\altaffilmark{12,13}}
\author{Cameron B. Hummels\altaffilmark{7}}
\author{Francesca Iannuzzi\altaffilmark{14}}
\author{Du\u{s}an Kere\u{s}\altaffilmark{15}} 
\author{Anatoly Klypin\altaffilmark{16}}
\author{Andrey V. Kravtsov\altaffilmark{3,10}}
\author{Mark R. Krumholz\altaffilmark{1}}
\author{Michael Kuhlen\altaffilmark{1,13}}
\author{Samuel N. Leitner\altaffilmark{17}}
\author{Piero Madau\altaffilmark{1}}
\author{Lucio Mayer\altaffilmark{18}} 
\author{Christopher E. Moody\altaffilmark{1}}
\author{Kentaro Nagamine\altaffilmark{19,20}}
\author{Michael L. Norman\altaffilmark{15}} 
\author{Jose O\~{n}orbe\altaffilmark{21}}
\author{Brian W. O'Shea\altaffilmark{22}}
\author{Annalisa Pillepich\altaffilmark{1}}
\author{Joel R. Primack\altaffilmark{23}}
\author{Thomas Quinn\altaffilmark{24}}
\author{Justin I. Read\altaffilmark{4}}
\author{Brant E. Robertson\altaffilmark{7}}
\author{Miguel Rocha\altaffilmark{21}}
\author{Douglas H. Rudd\altaffilmark{10, 25}}
\author{Sijing Shen\altaffilmark{1}}
\author{Britton D. Smith\altaffilmark{22}}
\author{Alexander S. Szalay\altaffilmark{26}}
\author{Romain Teyssier\altaffilmark{18}}
\author{Robert Thompson\altaffilmark{7, 19}}
\author{Keita Todoroki\altaffilmark{19}}
\author{Matthew J. Turk\altaffilmark{5}} 
\author{James W. Wadsley\altaffilmark{27}}
\author{John H. Wise\altaffilmark{28}}
\author{Adi Zolotov\altaffilmark{8} for the {\it AGORA} Collaboration\altaffilmark{29}}

\altaffiltext{1}{Department of Astronomy and Astrophysics, University of California at Santa Cruz, Santa Cruz, CA 95064, USA, {\tt me@jihoonkim.org}}
\altaffiltext{2}{Department of Physics, Stanford University, Stanford, CA 94305, USA}
\altaffiltext{3}{Department of Astronomy and Astrophysics, University of Chicago, Chicago, IL 60637, USA}
\altaffiltext{4}{Department of Physics, University of Surrey, Guildford, Surrey, GU2 7XH, United Kingdom}
\altaffiltext{5}{Department of Astronomy and Astrophysics, Columbia University, New York, NY 10027, USA}
\altaffiltext{6}{Department of Theoretical Physics, Universidad Autonoma de Madrid, Madrid, 28049, Spain}
\altaffiltext{7}{Department of Astronomy, University of Arizona, Tucson, AZ 85721, USA}
\altaffiltext{8}{Center for Astrophysics and Planetary Science, Racah Institute of Physics, The Hebrew University, Jerusalem, 91904, Israel}
\altaffiltext{9}{Particle Astrophysics Center, Fermi National Accelerator Laboratory, Batavia, IL 60510, USA} 
\altaffiltext{10}{Kavli Institute for Cosmological Physics, University of Chicago, Chicago, IL 60637, USA}
\altaffiltext{11}{Institute for Astronomy, ETH Zurich, Zurich, 8093, Switzerland}
\altaffiltext{12}{Department of Astronomy, California Institute of Technology, Pasadena, CA 91125, USA}
\altaffiltext{13}{Department of Astronomy, University of California at Berkeley, Berkeley, CA 94720, USA}
\altaffiltext{14}{Max-Planck Institut f\"{u}r Astrophysik, D-85741 Garching, Germany}
\altaffiltext{15}{Department of Physics, University of California at San Diego, La Jolla, CA 92093, USA}
\altaffiltext{16}{Department of Astronomy, New Mexico State University, Las Cruces, NM 88001, USA}
\altaffiltext{17}{Department of Astronomy, University of Maryland, College Park, MD 20742, USA}
\altaffiltext{18}{Institute for Theoretical Physics, University of Zurich, Zurich, 8057, Switzerland} 
\altaffiltext{19}{Department of Physics and Astronomy, University of Nevada, Las Vegas, NV 89154, USA}
\altaffiltext{20}{Department of Earth and Space Science, Graduate School of Science, Osaka University, 1-1 Machikaneyama, Toyonaka, Osaka, 560-0043, Japan}
\altaffiltext{21}{Department of Physics and Astronomy, University of California at Irvine, Irvine, CA 92697, USA}
\altaffiltext{22}{Lyman Briggs College and Department of Physics and Astronomy, Michigan State University, Lansing, MI 48825, USA}
\altaffiltext{23}{Department of Physics, University of California at Santa Cruz, Santa Cruz, CA 95064, USA}
\altaffiltext{24}{Department of Astronomy, University of Washington, Seattle, WA, 98195, USA}
\altaffiltext{25}{Research Computing Center, University of Chicago, Chicago, IL 60637, USA}
\altaffiltext{26}{Department of Physics and Astronomy, Johns Hopkins University, Baltimore, MD 21218, USA}
\altaffiltext{27}{Department of Physics and Astronomy, McMaster University, Hamilton, ON L8S 4M1, Canada}
\altaffiltext{28}{School of Physics, Georgia Institute of Technology, Atlanta, GA 30332, USA}
\altaffiltext{29}{The project website is http://www.AGORAsimulations.org/.}

\begin{abstract}
We introduce the {\it AGORA} project, a comprehensive numerical study of well-resolved galaxies within the $\Lambda$CDM cosmology.
Cosmological hydrodynamic simulations with force resolutions of $\sim 100$ proper pc or better will be run with a variety of code platforms to 
follow the hierarchical growth, star formation history, morphological transformation, and the cycle of baryons in and out of 8 galaxies with halo masses 
$M_{\rm vir}\simeq 10^{10}$, $10^{11}$, $10^{12}$, and $10^{13}\,\msun$ at $z=0$ and two different (``violent" and ``quiescent") assembly histories. 
The numerical techniques and implementations used in this project include the smoothed particle hydrodynamics codes {\sc Gadget} and {\sc Gasoline}, 
and the adaptive mesh refinement codes {\sc Art}, {\sc Enzo}, and {\sc Ramses}. The codes will share common initial 
conditions and common astrophysics packages including UV background, metal-dependent radiative cooling, metal and energy yields of supernovae, and stellar initial mass function. 
These are described in detail in the present paper.
Subgrid star formation and feedback prescriptions will be tuned to provide a realistic interstellar and circumgalactic medium 
using a non-cosmological disk galaxy simulation. Cosmological runs will be systematically compared with each other using a common analysis toolkit, and validated against observations to verify that the solutions are robust -- i.e., that the astrophysical assumptions are 
responsible for any success, rather than artifacts of particular implementations.  The goals of the {\it AGORA} project are, broadly speaking, to raise the 
realism and predictive power of galaxy simulations and the understanding of the feedback processes that regulate galaxy ``metabolism." The initial conditions for the {\it 
AGORA} galaxies as well as simulation outputs at various epochs will be made publicly available to the community.
The proof-of-concept dark matter-only test of the formation of a galactic halo with a $z=0$ mass of $M_{\rm vir} \simeq 1.7 \times 10^{11}\,\msun$ by 9 different versions of the participating codes is also presented to validate the infrastructure of the project.  
\end{abstract}

\section{Introduction}\label{intro}

\subsection{The State of Galaxy Simulation Studies}\label{state}

Ever since the dawn of high-performance computing, cosmological simulations have been the main theoretical tool for studying the hierarchical assembly of dark matter halos, the survival of substructure 
and baryon dissipation within the cold dark matter hierarchy, the flows of gas into and out of galaxies, and the nature of the sources responsible for the
reionization, reheating, and chemical enrichment of the Universe.  
Purely gravitational simulations of the distribution of dark matter on large scales using different codes -- e.g., {\it Millennium I} and {\it II} using {\sc Gadget} 
 \citep{MillenniumI, MillenniumII} and {\it Bolshoi} and {\it BigBolshoi/MultiDark} using {\sc Art} \citep{Bolshoi, BigBolshoi} -- now produce consistent results 
\citep{Springel12, Kuhlen12}. This is also true of collisionless ``zoom-in" high-resolution simulations: {\it Via Lactea II} \citep[using {\sc Pkdgrav-2};][]{VL2}, {\it Aquarius} 
\citep[using {\sc Gadget};][]{Aquarius}, and {\it GHALO} \citep[using {\sc Pkdgrav-2};][]{Ghalo}. To follow the formation and evolution of galaxies and clusters, however, it is necessary to model  baryonic physics, 
dissipation, chemical enrichment, the heating and cooling of gas, the formation of stars and supermassive black holes (SMBHs), magnetic fields, non-thermal plasma processes, along with the effects of the 
energy outputs from these processes.  A number of numerical techniques have been developed to treat gasdynamical processes in cosmological simulations,     
including Lagrangian smoothed particle hydrodynamics \citep[SPH;][]{GingoldMonaghan1977, 1977AJ.....82.1013L, Monaghan1992} and Eulerian adaptive mesh refinement \citep[AMR;][]{1984JCoPh..53..484B, BergerAMR}. Because of the complexity of the problem, the nonlinear 
nature of gravitational clustering, the different assumptions made regarding the cooling and heating functions of enriched, photoionized gas,  
and the different implementations of crucial gas subgrid physics, it is non-trivial to validate the results of different 
techniques and codes even when applied to similar astrophysical problems.

The {\it Santa Barbara Cluster Comparison} project \citep{SBCluster} first showed the benefit of comparing hydrodynamic simulations of the same astrophysical
system, a galaxy cluster, starting from the same initial conditions and using a large variety of codes. By modern standards, it was a 
relatively simple effort, in which the gas was assumed to be non-radiative. The spatial resolutions at the centers of simulations ranged from 5 to 400 kpc, and
outputs were compared at various redshifts.  The resulting simulations showed good agreement in the dark matter and gas density 
profiles, with a spread of about a factor of 2 in the predicted, resolution-dependent X-ray luminosity. Systematic differences were noted between 
SPH and grid-based codes including a mismatch in the central gas entropy profile, although 
the issue is now considered resolved \citep[e.g.,][]{2008MNRAS.387..427W, 2013arXiv1307.0668P}. 

One of the recent comparisons of hydrodynamic galaxy simulations is the {\it Aquila} project \citep{Aquila}, in which 13 different simulations were run from 
the same initial conditions using various implementations of the {\sc Gadget-3} \citep[an updated version of {\sc Gadget-2};][]{gadget2} and {\sc 
Gasoline} \citep{gasoline} SPH codes, the {\sc Arepo} moving mesh code \citep{arepo}, and the {\sc Ramses} AMR code \citep{ramses}. 
The initial conditions were those of the {\it Aquarius} halo {\it Aq-C}, with a mass at $z=0$ of $\sim 1.6\times 10^{12}\,\msun$, comparable to that of the 
Milky Way, and a relatively quiescent formation history.  
All groups participating in the comparison adopted their preferred implementation 
of radiative cooling, star formation, and feedback, and each simulation was run at two different resolutions to test convergence.
None of the {\it Aquila} simulations produced a disk galaxy resembling the Milky Way. Most runs resulted in unrealistic systems with too large a  
stellar mass and too little cold gas, a massive bulge and a declining rotation curve.  The stellar mass ranged from $\sim 4\times 10^{10}$ to $\sim 3\times 10^{11}\,\msun$.  Simulations with greater feedback led to lower stellar masses, but usually had a hard time producing a galactic disk with a rotation velocity in agreement with the Tully-Fisher relation of late-type spirals.
The star formation typically peaked at $z\sim4$ and declined thereafter, with essentially all simulations forming more than half their stars in the first $\sim 3$ Gyr.  The gaseous disk sizes were in better agreement with observations, but with too little gas mass. 
That the choice of numerical technique affected the {\it Aquila} results was shown, for example, by comparing {\sc Gadget-3} and {\sc Arepo} runs with the similar subgrid physics implementations: the 
{\sc Arepo} simulation produced almost twice as much stellar mass as the {\sc Gadget-3} simulation. The {\it Aquila} project shows the need to control the baryon 
overcooling, prevent the early burst of star formation \cite[e.g.,][]{2000MNRAS.315L..18E}, and promote the accretion and retention of late-accreting high-angular-momentum baryons needed to form spiral disks.

\begin{table*}
\caption{Task-oriented Working Groups of the {\it AGORA} Project}
\centering
\begin{tabular}{c || c}
\hline\hline 
Working Group & Objectives and Tasks\tablenotemark{{\it a}} \\ 
\hline
{\it Common Cosmological ICs} & Determine common initial conditions for cosmological high-resolution zoom-in galaxies (Section \ref{CosICs})  \\
{\it Common Isolated ICs} & Determine common initial conditions for an isolated low-redshift disk galaxy (Section \ref{DiskICs})  \\
{\it Common Astrophysics} & Define common physics including UV background, gas cooling, stellar IMF, energy and metal yields from SNe (Section \ref{physics}) \\ 
{\it Common Analysis} & Support common analysis tools, define physical and quantitative  comparisons across all codes  (Section \ref{analysis}) \\
\hline
\end{tabular}
\tablenotetext{1}{\scriptsize For detailed explanation, see the referenced section.}
\label{table:task-oriented}
\end{table*}

On the other hand, a new cosmological simulation of extreme dynamic range, {\it Eris} \citep{Eris}, succeeded for the first time to produce a realistic massive late-type galaxy at $z=0$ in which the structural properties, the mass budget in the various components (disk, bulge, halo), and the scaling relations between mass and luminosity 
are all consistent with a host of observational constraints. Run with the {\sc Gasoline} code, {\it Eris} had 25 times better mass resolution than the typical  
{\it Aquila} simulation, and adopted a blastwave scheme for supernova feedback \citep{stinson06} that generates galactic outflows without explicit wind particles.\footnote{The force softening of the {\it Aquila} {\sc Gasoline} simulation was 460 proper pc  from $z=9$ to $z=0$, and the mass per dark matter and gas particle was 2.1 and $0.5\times 10^6 
\,\msun$, respectively.  The {\it Eris} simulation had the force softening of 120 proper pc from $z=9$ to $z=0$, and better mass per dark matter and gas particle of 9.8 and $2.0\times 10^4 \msun$, respectively.}
Combined with a high gas density threshold for star formation, this scheme has been found to be key also in producing realistic dwarf galaxies \citep{governato10}. 
Indeed, the importance of reaching a high star formation threshold has been well studied since it was first pointed out by \cite{Kravtsov2003}. 
It enables energy deposition by supernovae within small volumes, and the development of an inhomogeneous interstellar medium where star formation and heating by supernovae occur in a clustered fashion. 
The resulting outflows at high redshifts reduce the baryonic content of galaxies and preferentially
remove low angular momentum gas, decreasing the mass of the bulge component \citep{2011MNRAS.415.1051B}.
All in all, high numerical resolution appears to be essential if simulations are to resolve the regions where stars form, and thus to succeed in producing realistic galaxies. 
Yet, even at {\it Eris}'s resolution one barely resolves the vertical disk structure of the Milky Way since the Milky Way H~\textsc{i} disk scale height is about 120 pc \citep{Lockman84} and the ${\rm H}_2$ scale height is about 50 pc inside the solar circle \citep{Sanders84, NarayanJog02}. 

\subsection{Need for High-Resolution Galaxy Simulation}\label{need}

As the above discussion should make clear, the success of cosmological galaxy formation simulations in producing realistic galaxies is a strong function of resolution (and the corresponding subgrid physics; e.g., higher star formation threshold). 
Numerically resolving the star-forming regions and the disk scale height is necessary because the interplay between the simulation resolution and the realism of subgrid models of star formation and feedback processes is increasingly thought of as a key to successful modeling of galaxy formation \citep[e.g.,][]{2011MNRAS.410.1391A, 2012ApJ...749..140H}.
In retrospect this is not surprising.  Stars form, and deposit at least some of their feedback, in the densest, coldest phase of the interstellar medium (ISM). The characteristic property of star-forming regions is that they have extinctions high enough to block out the ultraviolet (UV) starlight that pervades most interstellar space. In the absence of UV light, the ISM undergoes a phase transition from H~\textsc{i} to H$_2$ and the gas temperature drops to $\sim 10\,$K, which is likely the critical step in the onset of star formation \citep{krumholz11b, glover12a}. In the Local Group, the characteristic sizes of these star-forming molecular clouds are only $\sim 10-100$ pc, and they occupy a negligible fraction of the ISM volume, but contain a non-trivial fraction of its total mass: $\sim 30\%$ of in the Milky Way, with lower fractions in dwarf galaxies and higher fractions in larger galaxies with denser interstellar media \citep{blitz07a}. Despite molecular clouds' high density, however, star formation within them remains surprisingly inefficient. In nearby galaxies the observed star formation timescales in molecular gas is $\sim 2$ Gyr \citep[e.g.,][]{Bigiel2008, Schruba11a}, and it has been known for more than 30 years that, on average, the molecular gas forms stars at a rate of no more than $\sim 1\%$ of the mass per dynamical time \citep{zuckerman74a, krumholz07e, krumholz12a}. It is clear that any successful model for galaxy formation must include an adequate model for this critical, high-density phase, hence the need for high resolution.
High resolution is also essential to avoid numerical loss of angular momentum for SPH codes that may alter the  kinematics and morphology of the disk and spheroid, and may lead to overly massive bulges and steep rotation curves \citep[e.g.,][]{2007MNRAS.375...53K, 2008ASL.....1....7M}. 

The simulations in our project seek to follow the processes that regulate star formation on small scales as faithfully as possible. There are a number of feedback processes whose relative importance likely depends on the type of galaxy and the spatial scale that one is considering (see \citealt{dekel13a} for a recent summary), including photoionization that heats gas to $\sim 10^4$ K and disperses star-forming clouds \citep{whitworth79a, matzner02a, dale12a}, fast stellar winds that shock-heat the ISM and produce expanding bubbles \citep{castor75a, weaver77a, chevalier85a}, the pressure of both direct starlight and dust-reprocessed radiation \citep{krumholz09d, murray10a, 2011MNRAS.417..950H, 2012ApJ...760..155K, 2013arXiv1302.4440K}, and energy injection by Type Ia and Type II supernovae. It is clear from both observations \citep[e.g.,][]{lopez11a} and theory \citep[e.g.,][]{2011MNRAS.417..950H, 2013MNRAS.428..129S} that these feedback processes interact with one another in non-trivial ways. For example, ionization and momentum from radiation pressure and stellar winds act on short time and spatial scales when massive stars are formed to ``clear out'' the dense regions of the GMC in which they form, and ionize and heat the surrounding neighborhood. This allows hot gas from shocked supernovae ejecta -- which occur several Myrs later (often after the cloud is mostly destroyed) -- to escape and couple to the more diffuse ISM, preventing its rapid cooling and allowing it to expand further and drive outflows.

By following these processes as directly as possible, and constraining against well-tested simulations of local systems and the ISM, one of the goals of the {\it AGORA} project is to lift the degeneracies between the subgrid treatments of current cosmological galaxy formation models. Indeed, the largest barrier to using today's cosmological simulations to constrain fundamental physics of cooling, shocks, turbulence, the ISM, star formation, and dark matter on sub-galactic scales is probably the unconstrained degrees of freedom in subgrid treatments of the ISM. At the same time, simulations of star formation and feedback in isolated galaxies or sub-regions of them have for the most part been run only over a narrow range in redshift, metallicity, and structural properties, without the context provided by realistic dark matter halos or baryonic inflows. The simulations to be studied in the present project provide an opportunity to explore the physics of star formation regulation in a fully cosmological context. Only by iterating between small-scale, high-resolution simulations and cosmological ones like {\it AGORA} can we hope to reach a complete theory of galaxy formation.

\begin{table*}
\caption{Science-oriented Working Groups of the {\it AGORA} Project}
\centering
\begin{tabular}{c||c}
\hline\hline 
Working Group & Science Questions (includes, but are not limited to)\tablenotemark{{\it a}} \\ 
\hline
{\it Isolated Galaxies and Subgrid Physics} & Tune subgrid models across codes to yield similar results for similar astrophysical assumptions \\ 
{\it Dwarf Galaxies} & Simulate cosmological $\sim10^{10}\,\msun$ halos and compare results across all participating codes  \\
{\it Dark Matter} & Radial profile, shape, substructure, core-cusp problem  \\
{\it Satellite Galaxies}  & Effects of environment, UV background, tidal disruption  \\
{\it Galactic Characteristics} & Surface brightness, disks, bulges, metallicity, images, spectral energy distributions  \\
{\it Outflows} & Galactic outflows, circumgalactic medium, metal absorption systems  \\
{\it High-redshift Galaxies} & Cold flows, clumpiness, kinematics, Lyman-limit systems  \\
{\it Interstellar Medium} & Galactic ISM, thermodynamics, kinematics  \\
{\it Massive Black Holes} & Growth and feedback of massive black holes in a galactic context  \\
\hline
\end{tabular}
\tablenotetext{1}{\scriptsize See Section \ref{WGs} and the project website for more information on the working groups.}
\label{table:science-oriented}
\end{table*}

\subsection{Motivation and Introduction of the {\it AGORA} Project}\label{motivation}

This motivated us to organize a new galaxy ``zoom-in" simulations comparison project, with emphases on resolution, the physics of the ISM, feedback and galactic 
outflows, and initial conditions covering a range of halo masses from dwarfs to massive ellipticals. 
We also wanted to require more similarity in the astrophysical assumptions used in the simulations.
A meeting was held at the University of California, Santa Cruz, in August 2012 to initiate this project. As of this writing, 95 individuals from 
47 different institutions worldwide, using a variety of platform codes, have now agreed to participate in what has been named the {\it Assembling Galaxies of 
Resolved Anatomy} ({\it AGORA}) Project.\footnote{See the project website at http://www.AGORAsimulations.org/ for further information on the project, 
and its membership and leadership.}  

In this paper we describe the goals, infrastructures, techniques, and tools of the {\it AGORA} simulations comparison
project. These should be of interest not only to the groups participating in {\it AGORA} but also to other groups conducting galaxy simulations, since one of the 
purposes of the project is to increase the level of realism in all such simulations. The numerical techniques and implementations used in this project 
include the smoothed particle hydrodynamics codes {\sc Gadget} and {\sc Gasoline},
and the adaptive mesh refinement codes {\sc Art}, {\sc Enzo}, and {\sc Ramses} (see Section \ref{proof-of-concept-codes} for more information).  
These codes will share common initial conditions (Section \ref{ICs}), and common astrophysics packages (Section \ref{physics}) including photoionizing UV background, metal-dependent radiative cooling, metal and energy yields, stellar initial mass function, and will be systematically compared with each other and against a variety of observations using a common analysis toolkit (Section \ref{analysis}).  
The goals of the {\it AGORA} project are, broadly speaking, to raise the realism and predictive power of galaxy simulations and the understanding of the feedback processes that regulate galaxy ``metabolism", and by doing so to solve long-standing problems in galaxy formation.  

In order to achieve this goal the {\it AGORA} project will employ simulations designed with state-of-the-art resolution.
Since it is clear that the interplay between resolution 
and subgrid modeling of star formation and feedback is one of the key aspects of modeling galaxy formation our choice
is mandatory to make progress in this area despite the expected large computational cost.
Doing this in a variety of code platforms is also essential, not only for the benefit of the groups using each code, but also {\it to verify that the solutions are robust} -- i.e., that the astrophysical assumptions are responsible for any success, rather than artifacts of particular implementations.  
This way, the project will enable improved scientific understanding of galaxy formation, a key subject that is at last yielding to a combination of theory and computation.
We plan to achieve this by sharing outputs at many redshifts, with many groups analyzing each issue using common analysis code applied to outputs from multiple codes -- even timestep by timestep in some cases.  Many of these intermediate timesteps along with the common initial conditions will be made available to the community.  

To build the infrastructure of the {\it AGORA} project four task-oriented working groups have been established (see Table \ref{table:task-oriented}).  These working 
groups ensure that the comparison of simulations is bookended by common initial conditions, common astrophysical assumptions, and common analysis tools. 
We also have initiated nine science-oriented working groups, each of which aims to perform original research (see Table \ref{table:science-oriented}; see also the {\it AGORA} project website for leaderships and memberships of these groups) and 
address basic problems in galaxy formation both theoretically and observationally.  In other words, the {\it AGORA} project is not just a single set of 
simulations being compared, but {\it a launchpad to initiate a series of science-oriented subprojects}, each of which is independently designed, executed, 
and studied by members of the science-oriented working groups. 

\begin{table*}
\caption{A Suite of Cosmological Initial Conditions\tablenotemark{{\it a}}}
\centering
\begin{tabular}{c || c | c | c | c}
\hline\hline 
 & Isolated Dwarfs & Sub-$L_{\star}$ Galaxies & Milky Way-sized Galaxies & Ellipticals or Galaxy Groups \\ 
\hline
Halo virial mass at $z=0$ & $\sim 10^{10} \msun$ & $\sim 10^{11} \msun$ & $\sim 10^{12} \msun$ & $\sim 10^{13} \msun$\\
Maximum circular velocity & $\sim 30 \,\,{\rm km\,s^{-1}}$ & $\sim 90 \,\,{\rm km\,s^{-1}}$ &  $\sim 160 \,\,{\rm km\,s^{-1}}$ & $\sim 250 \,\,{\rm km\,s^{-1}}$ \\
Selected merger histories & quiescent/violent at $z<2$ & quiescent/violent at $z<2$ & quiescent/violent at $z<2$ & quiescent/violent in $2<z<4$ \\
\hline
\end{tabular}
\tablenotetext{1}{\scriptsize For detailed explanation, see Section \ref{CosICs}.}
\label{table:CosICs}
\end{table*}

An example of a problem in galaxy formation that we want to address is the mechanisms that lead to galactic transformations -- the processes that form galactic spheroids from disks (e.g., mergers versus disk instability), and the processes that quench star formation in galaxies (e.g., AGN feedback versus cutoff of cold flows as halo mass increases). 
An important constraint that has emerged in the last few years, halo abundance matching \citep[e.g.,][]{Behroozi12, Moster13},  has led to a ``stellar mass problem'', namely, for a given halo mass, the combined mass of stellar disks and stellar bulges is too large in numerical simulations of galaxy formation relative to the expectations.  Only a handful of simulations have been shown to be consistent with such constraint at $z=0$ including {\it Eris} \cite[e.g.,][]{2013ApJ...766...56M}, but all are seriously discrepant at higher redshift.
Another difficult problem in galaxy formation is the mutual effects of baryons on dark matter, and dark matter on baryons, in accounting for the radial distribution, kinematics, and angular momentum of stars and gas in observed galaxies.  It appears that compression of the central dark matter due to baryonic infall \citep[e.g.,][]{BFFP86,Gnedin04,Gnedin11} is an important effect in early-type galaxies, but this is apparently largely offset by other effects in late-type galaxies \citep[e.g.,][]{Dutton07,T-G,Dutton12}.  Thus important astrophysical phenomena appear to cause opposite effects, which makes this a challenging problem.  
In order to address these processes of galaxy formation, it will be necessary to understand better the astrophysics of star formation and feedback, and the fueling and feedback from SMBHs -- both of which are treated as subgrid physics in galaxy-scale simulations  \citep[e.g.,][]{2011ApJ...738...54K}.  We will tackle this challenge by carefully comparing simulations using different codes and different subgrid implementations with each other and with observations, and also by simultaneously improving the theoretical understanding of these processes, including running very high-resolution simulations on small scales.

The remainder of this paper is organized as follows.  
Section \ref{ICs} discusses the initial conditions for the {\it AGORA} simulations, both cosmological and isolated.  Two sets of cosmological initial conditions are generated using the MUlti-Scale Initial Conditions code \citep[{\sc Music};][]{MUSIC} for halos with masses at $z=0$ of about $10^{10}$, $10^{11}$, $10^{12}$, $10^{13}\,\msun$, one set with a quiescent merger history and the other set with many mergers.  Additional initial conditions are generated for an isolated disk galaxy with gas fraction and structural properties characteristic of galaxies at $z\sim1$, to which the same simulation codes will be applied. 
Section \ref{physics} discusses the common astrophysical assumptions to be applied in all of the hydrodynamic simulations, such as the UV background, the metallicity-dependent gas cooling, and the stellar initial mass function (IMF) and metal production assumptions. 
Section \ref{simulations} describes strategies for running the simulations and comparing them at many redshifts, with each other and with observations. The {\tt yt} analysis code \citep{yt} will be instrumental in comparing the simulation outputs, as it takes as input the outputs from all of the simulation codes being studied.  The objectives of science-oriented comparison of the simulation outputs are discussed, too. 
Section \ref{proof-of-concept} demonstrates the proof-of-concept dark matter-only simulation for a galactic halo with a $z=0$ mass of $M_{\rm vir} \simeq 1.7 \times 10^{11}\,\msun$ to field-test the pipeline of the project, including the common initial conditions and common analysis platform.  
The new results and ideas established in this paper are summarized in Section \ref{conclusion}.

\section{Common Initial Conditions}\label{ICs}

In this section the common initial conditions to be employed in the {\it AGORA} simulations are introduced, both cosmological and isolated.  A companion paper in preparation \citep{IC-paper} will further present the initial conditions of the project in more detail.  We also note that the {\sc Music} parameter files to generate cosmological initial conditions, as well as the initial conditions themselves in formats suitable for all of the simulation codes being used in the project, will be publicly available through the {\it AGORA} project website.  

\begin{table*}
\caption{Components of Isolated Disk Initial Conditions\tablenotemark{{\it a}}}
\centering
\begin{tabular}{c || c | c | c | c}
\hline\hline 
 & Dark Matter Halo & Stellar Disk & Gas Disk & Stellar Bulge \\ 
\hline
Density profile & \cite{1997ApJ...490..493N} & Exponential & Exponential & \cite{1990ApJ...356..359H}\\
\multirow{2}{*}{Physical properties}  & $v_{\rm c, 200} = 150 {\rm \,\,km\,s^{-1}}$, $M_{200} = 1.074\times10^{12} \msun$, & $M_{\rm d}=4.297\times 10^{10} \msun$, & \multirow{2}{*}{$f_{\rm gas}=20\%$} & \multirow{2}{*}{bulge-to-disk mass ratio $\,\,B/D=0.1$} \\
 & $r_{200}=205.4$ kpc, $c=10$, $\lambda=0.04$  & $r_{\rm d}=3.432$ kpc, $z_{\rm d}=0.1\,r_{\rm d}$ & & \\
Number of particles  & $10^5$ (low-res.), $10^6$ (medium), $10^7$ (high) & $10^5$, $10^6$, $10^7$ &  $10^5$, $10^6$, $10^7$ & $1.25\times10^4$, $1.25\times10^5$, $1.25\times10^6$ \\
\hline
\end{tabular}
\tablenotetext{1}{\scriptsize For detailed explanation and definition of the parameters, see Section \ref{DiskICs}.}
\label{table:DiskICs}
\end{table*}

\subsection{Cosmological Initial Conditions}\label{CosICs}

Common cosmological initial conditions for all simulation codes are generated using the {\sc Music} code \citep[MUlti-Scale Initial Conditions;][]{MUSIC}.\footnote{The website is http://bitbucket.org/ohahn/music/.} {\sc Music} uses a real-space convolution approach in conjunction with an adaptive multi-grid Poisson solver to generate highly accurate nested density, particle displacement and velocity fields suitable for multi-scale ``zoom-in'' simulations of cosmological structure formation. 
For the run with two-component baryon$+$cold dark matter (CDM) fluid, we assume in the {\it AGORA} project that density perturbations in both fluids are equal and follow the total matter density perturbations (i.e., CDM and baryon perturbations have the same power spectrum; good approximation at late times).
For particle-based codes, baryon particles are generated on a staggered lattice with respect to the CDM particles and displaced by the same displacement field as the CDM particles evaluated at the staggered positions. This strategy is to minimize two-body effects.  For grid-based codes, the density perturbations are generated directly on the mesh using the local Lagrangian approximation. In both cases, the initial temperature is set to the cosmic mean baryon temperature at the starting redshift.
We refer the reader to \cite{MUSIC} for details on the algorithms employed, but describe aspects that are particularly relevant for the {\it AGORA} project in what follows.  {\sc Music} generates the various fields on a range of nested levels $\ell$ of effective linear resolution $2^\ell$. The level covering the entire computational domain is called {\tt levelmin}, $\ell_{\rm min}$, and the maximum level {\tt levelmax}, $\ell_{\rm max}$.  

\begin{itemize}
\item {\it White noise generation and phase consistency:}  The white noise fields that source all perturbation fields are drawn reproducibly from a sequence of random number seeds  $\left\{s_i\right\}$ where $i \in \left[ \ell_{\rm min}, \ell_{\rm max} \right]$. 
The random number generator used in {\sc Music} is one that comes with the GNU Scientific Library, so the white noise fields are the identical on any machine as long as they are drawn from the same seeds $\left\{s_i\right\}$.
Specifying this sequence of numbers thus entirely defines the ``universe'' for which {\sc Music} generates the initial conditions. 
Particular ``zoom-in'' regions of high resolution can be shifted or enlarged, and the resolution can be increased or decreased, without breaking consistency.  
This means that the {\sc Music} parameter file can be distributed rather than a binary initial conditions file.   Appendix \ref{app-MUSIC} illustrates an example of such parameter files.  
\item {\it Multi-code compatibility:}  {\sc Music} supports initial conditions for baryons and dark matter particles for a wide range of cosmological simulation codes, many of which represented also in the {\it AGORA} project. Support for the various simulation codes and their various file formats for initial condition files is achieved through a {\tt C++} plugin mechanism.  This allows adding new output formats without touching any parts of the code itself, and code specific parameters can be added transparently.
\item {\it Expandability:}  The current set-up allows expandability in various respects: {\it 1)} Support for new simulation codes can be added through plugins rather than file conversion. {\it 2)} The size and resolution of the high-resolution region can be altered consistently. {\it 3)} Future simulations focusing on larger regions or higher resolution are easily possible, and will be consistent with existing simulations. {\it 4)} Cosmological models can be changed easily, and alternative perturbation transfer functions, e.g., for distinct baryon and DM perturbations or for warm dark matter, can be easily adopted. 
\end{itemize}

Using the {\sc Music} code, we have generated two sets of cosmological initial conditions for high-resolution zoom-in simulations targeted at halos with $z=0$ masses of about $10^{10}$, $10^{11}$, $10^{12}$, $10^{13}\,\msun$, one set with a {\it quiescent} merger history (i.e., relatively few major mergers) and the other set with a {\it violent} merger history (i.e., many mergers between $z=2$ and 0 for a $\sim 10^{10} - 10^{12} \,\msun$ halo).  
Physically, these choices cover from the formation of dwarf galaxies to elliptical galaxies and to galaxy groups (see Table \ref{table:CosICs}).  
First-order Lagrangian perturbation theory is used (default in the {\sc Music} code) to initialize displacements and velocities of dark matter particles.
The $\Lambda$CDM cosmological parameters we adopt are consistent with the  {\it WMAP} 7/9 results \citep{WMAP7, WMAP9} that includes additional cosmological data from ground-based observations of the Type Ia supernovae (SNe) and the baryonic acoustic oscillation (BAO): $\Omega_{\rm m}=0.272$, $\Omega_\Lambda=0.728$, $\sigma_8=0.807$, $n_{\rm s}=0.961$, and $H_0= 70.2\  {\rm km\ } {\rm s}^{-1} {\rm Mpc}^{-1}$.
Radiation energy density and the curvature terms are assumed to be negligible: $\Omega_{\rm R}=\Omega_{\rm k}=0$.  
By performing a comparison study we find that adopting the latest {\it Planck} cosmology \citep{2013arXiv1303.5076P} does not noticeably change the properties of individual halos.  
While the present paper employs {\it WMAP} cosmology, the {\it AGORA} Collaboration may later decide whether to switch to more recent cosmological parameters.

First, low-resolution dark matter-only pathfinder simulations are performed from $z=100$ to $z=0$ to identify halos of appropriate merger histories.  They are carried out in a $(5\, h^{-1}\,{\rm comoving \,\, Mpc})^3$ box for $\sim 10^{10} \msun$ halos, and in a $(60\, h^{-1}\,{\rm comoving \,\, Mpc})^3$ box for $\sim 10^{11} - 10^{13} \,\msun$ halos.  A strong isolation criterion is imposed for the {\it quiescent} set of the initial conditions to select target halos.  That is, the $3R_{\rm vir}$ radius circle of the halo being simulated must not intersect the $3R_{\rm vir}$ radius circle of any halo with half or more of its mass at $z=0$.  A relaxed criterion is used for the {\it violent} set of the initial conditions: $2R_{\rm vir}$ circle instead of $3R_{\rm vir}$.  Then, a higher-resolution dark matter-only simulation (e.g., particle resolution of $\sim 3\times10^5\,\msun$ for $\sim 10^{11} - 10^{13} \,\msun$ halos) is performed on a new initial condition re-centered on each of the target halos with nested resolution elements around it.  

The highest-resolution region in this initial condition is sufficiently large to include all the structures that merge with the target galaxy or have a significant impact on its evolution. 
The highest-resolution region is also carefully selected so that the target halo is ``contaminated" only by a minimal number of lower-resolution particles at final redshifts ($z=0$ for $\sim 10^{10}-10^{12}\,\msun$ halos; $z=2$ for $\sim 10^{13}\,\msun$ halos).
While this region is typically a superset of a Lagrangian volume of the target halo's $\sim 2R_{\rm vir}$ sphere at the final redshift, it should also be as small as possible in order to minimize the computational cost (i.e., cpu-hours, memory consumption).  
To this end, {\sc Music} supports highest-resolution particles to be placed only in a minimum bounding ellipsoid of the Lagrangian volume.\footnote{Using such an ellipsoidal initial condition for a Lagrangian region of a $2.5R_{\rm vir}$ radius sphere of a $M_{\rm vir} \simeq 1.7\times10^{11} \msun$ halo at $z=0$, the contamination level by lower-resolution particles inside $R_{\rm vir}$ is found to be $<$ 0.01\% in mass at $z=0$ (tested with $\left[ \ell_{\rm min}, \ell_{\rm max} \right] = [7, 12]$ on the {\sc Ramses} code.)}
For particle-based codes, this determines the position of the highest-resolution region directly, while for grid-based codes, an additional refinement mask is generated that traces the high-resolution region.
A high-resolution dark matter run is used to iteratively adjust the highest-resolution region by checking the contamination level inside the target halo at a final redshift.
Initial conditions generated this way have been verified readable by all the participating codes in our proof-of-concept tests (see Section \ref{proof-of-concept}).  

\subsection{Isolated Disk Initial Conditions}\label{DiskICs}

Stellar feedback processes are implemented quite differently in our different participating codes.  Modeling properly supernovae explosions or radiation from young stars remains a challenge when the target spatial resolution is 100 pc. Most, if not all, current feedback implementations are therefore highly phenomenological, and they are based on various parameters that need to be calibrated on required observational properties of simulated galaxies (e.g., star formation rate, gas and stellar fraction, H~\textsc{i} versus stellar mass, metallicity). Moreover, most of the proposed models depend strongly on the adopted mass and spatial resolution. It is of primary importance to understand how each individual code needs to be calibrated to reproduce various observational constraints. In a comparison like the {\it AGORA} project, it is even more important to cross-calibrate stellar feedback processes of the various codes using an idealized set-up such as an isolated disk. This is precisely the goal of this second type of initial condition: we would like to model a realistic galactic disk using our various codes and their feedback recipes, varying both the feedback parameters and the mass and spatial resolutions. 
By doing so, subgrid star formation and feedback prescriptions in various code platforms will be tuned to provide a realistic interstellar and circumgalactic medium. 
We will use for that a well-defined set of observables (e.g., star formation rate, stellar and cold gas fractions, metal-enriched circumgalactic medium) to identify {\it for each code} the appropriate set of feedback parameters as a function of the mass resolution. It is only after this first careful step that we will be able to move towards our final goal, namely comparing codes in cosmological simulations. 

The isolated disk galaxy initial conditions with gas fraction and structural properties characteristic of galaxies at $z\sim1$ are generated using the {\sc MakeDisk} code written by Volker Springel.
This code is based on solving the Jeans equations for a quasi-equilibrium multi-component halo/disk/bulge collisionless system, the particle distribution function in velocity space being assumed to be Maxwellian. ÊOur initial conditions  are quasi-equilibrium 4-component systems: the dark matter halo of a circular velocity of $v_{\rm c, \,200} = 150 {\rm \,\,km\,s^{-1}}$ has a mass of $M_{200} = 1.074\times10^{12} \msun$, and follows \citet[][NFW]{1997ApJ...490..493N} profile with corresponding concentration parameter $c=10$ and spin parameter $\lambda=0.04$. The disk follows an exponential profile (as a function of the cylindrical radius $r$ and the vertical coordinate $z$) with scale length $r_{\rm d}=3.432$ kpc and scale height $z_{\rm d}=0.1\,r_{\rm d}$. The disk is decomposed into a stellar component of mass $M_{\rm d}=4.297\times 10^{10} \, \msun$ and a gaseous component with $f_{\rm gas}=M_{\rm d, \,gas}/M_{\rm d}=20\%$. The last ingredient is a stellar bulge that follows the \cite{1990ApJ...356..359H} density profile with bulge-to-disk mass ratio $B/D=0.1$.  

We have generated 3 different sets of initial conditions that differ in the number of resolution elements used to describe each component  (see Table \ref{table:DiskICs}).
The low-resolution disk has $10^5$ particles for the halo, the stellar disk and the gaseous disk, and only $1.25\times 10^4$ particles for the bulge. The medium resolution version has 10 times more particles in each component, and the high-resolution one has 100 times more elements. Note that for the gas disk, we provide particle data that can be used directly by SPH or indirectly by grid-based codes via projecting the particles into a grid. On the other hand, the default option for grid-based codes is to use the analytical density profile for the gaseous component which is just
\begin{equation}
\rho(r,\,z)=\rho_0e^{-r/r_{\rm d}}\cdot e^{-|z|/z_{\rm d}}
\end{equation}
with $\rho_0= M_{\rm d} / 4\pi r_{\rm d}^2 z_{\rm d}$.
All codes will use $10^4$~K for the initial disk temperature.  In grid-based codes, we also need to set the gas density, pressure and velocity in the halo. We recommend to use for the halo zero velocity, zero metallicity, low gas density $n_{\rm H}=10^{-7}$~${\rm  cm}^{-3}$ and high gas temperature of $10^6$~K. This way, the total mass of gas coming from Êthe halo is negligible compared the gas disk mass.
We refer interested readers to an upcoming article in preparation for more details on the isolated disk initial conditions (e.g., the required spatial resolution for each mass resolution) and the results from the test using such initial conditions.  

\begin{figure*}
    \centering
    \includegraphics[width=0.77\textwidth]{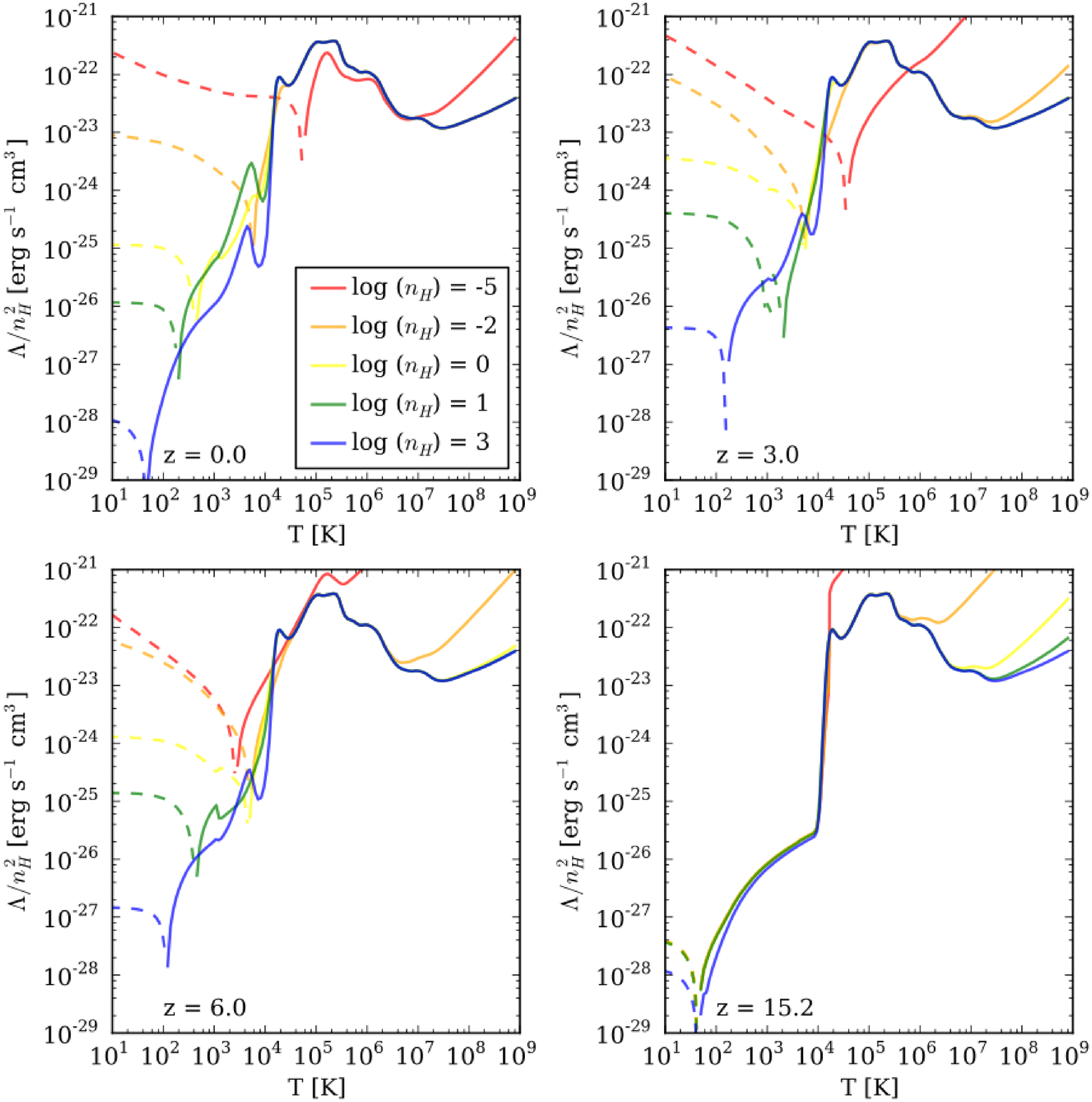}
      \caption{Gas cooling in the {\it AGORA} simulations. Equilibrium cooling rates normalized by $n_{\rm H}^2$ calculated with 
      the {\sc Grackle} cooling library for H number densities of 10$^{-5}$ ({\it red}), 10$^{-2}$ ({\it orange}), 1 ({\it yellow}), 10 ({\it green}), 
      and 10$^{3}$ ({\it blue}) cm$^{-3}$ at redshifts $z=0, 3, 6,$ and 15.2 (just before the UV background turns on) and solar metallicity gas.  
      {\it Solid lines} denote net cooling and {\it dashed lines} denote net heating.  The curves plotted are made with the non-equilibrium chemistry network of H, 
      He, H$_{2}$, and HD with tabulated metal cooling assuming the presence of a UV metagalactic background from \citet{HaardtMadau12}. 
\label{fig:cooling}}
     \vspace{0.4cm}
\end{figure*}

\section{Common Astrophysics}\label{physics}

We describe in this section the common astrophysics packages adopted by default in all {\it AGORA} simulations.  They include the metallicity-dependent gas cooling, UV background, stellar IMF, star formation, metal and energy yields by supernovae, and stellar mass loss.  

\subsection{Gas Cooling}\label{cooling}

The rate at which diffuse gas cools radiatively determines the response of baryons to dark matter potential wells, regulates star formation, controls
stellar feedback, and governs the interaction between galactic outflows and the circumgalactic medium (CGM). The ejection of the nucleosynthetic products of
star formation into the ISM modifies its thermal and ionization state, as radiative line transitions of carbon, oxygen, neon, and iron
significantly reduces the cooling time of enriched gas in the temperature range $10-10^7$ K. The picture is further complicated
by the presence of ionizing radiation, which removes electrons that would otherwise be collisionally excited and reduces the net
cooling rates. Photoionization increases the relative importance of oxygen as coolant, and decreases that of carbon, helium, and especially hydrogen
\citep{Wiersma09}. Both ionizing background radiation and metal line cooling must be included for the cooling rates to be correct to within a few
orders of magnitude \citep{2011MNRAS.413..190T}.

All {\it AGORA} simulations will use a standardized chemistry and cooling library, {\sc Grackle}.\footnote{The website is
http://grackle.readthedocs.org/.}\,  {\sc Grackle} provides a non-equilibrium primordial chemistry network for atomic H and He \citep{1997NewA....2..181A,
1997NewA....2..209A}, H$_2$ and HD \citep{2002Sci...295...93A, 2009Sci...325..601T}, Compton cooling off the cosmic microwave background (CMB), tabulated metal cooling and photo-heating rates calculated
with the photo-ionization code {\sc Cloudy} \citep{2013RMxAA..49..137F}.\footnote{The website is http://www.nublado.org/.} 
{\sc Grackle} also provides a look-up table for equilibrium cooling; depending on the problem at hand, both solvers may be used by the {\it AGORA} simulations.
At each
redshift, the gas is exposed to the CMB radiation and a uniform ultraviolet background (UVB) and is assumed to be dust-free and optically thin.
The metals are assumed to be in ionization equilibrium, so one can calculate in advance the cooling rate for a parcel of gas with a given density,
temperature, and metallicity, that is photoionized by incident radiation of known spectral shape and intensity. Following  \cite{Kravtsov2003}, \citet{Smith08}, \citet{2008ApJ...680.1083R}, \citet{Wiersma09},
and \citet{Shen10}, we will use pre-computed tabulated rates from the photoionization code {\sc Cloudy} at all temperatures in the
range $10-10^9$ K (see Figure \ref{fig:cooling}). {\sc Cloudy} calculates an equilibrium solution by balancing the incident heating with the radiative cooling from a full
complement of atomic and molecular transitions up to atomic number 30 (Zn). All metal cooling rates are tabulated for solar abundances as a 
function of total hydrogen number density, gas temperature, and redshift (as the radiation background evolves with time, see below),
and are scaled linearly with metallicity.  Instead of allowing {\sc Cloudy} to cycle through temperatures until converging on a
thermodynamic equilibrium solution, we will use the ``constant temperature" command to fix the temperature externally, allowing us to utilize 
{\sc Cloudy}'s sophisticated machinery to calculate cooling rates out of thermal equilibrium.  We will also deactivate the H$_{2}$ chemistry
in {\sc Cloudy} with the ``no H2 molecule" command since it is solved directly by the non-equilibrium chemistry solver in {\sc Grackle}.  
Because we directly solve for the electron density and the ionization of the most abundant elements, we are able to  
calculate the mean molecular weight and the gas temperature.

\subsection{Star Formation Prescription}\label{SFR}

The default {\it AGORA} simulation will follow only the atomic gas phase, with star formation proceeding at a rate 
\begin{equation}
{d\rho_* \over dt} = {\epsilon \rho_{\rm gas} \over t_{\rm ff}} \propto \rho_{\rm gas}^{1.5}
\label{eq:KS}
\end{equation}
(i.e., locally enforcing the Schmidt law), where $\rho_*$ and $\rho_{\rm gas}$ are the stellar and gas densities, $\epsilon$ is the star formation efficiency, and $t_{\rm ff}=\sqrt{3\pi/(32 G\rho_{\rm gas})}$ is the local free-fall time. 
We will also apply a density threshold below which star formation is not allowed to occur, and a non-thermal pressure floor to stabilize scales of order the smoothing length or the grid cell against gravitational collapse and avoid artificial fragmentation \citep{1997MNRAS.288.1060B, 1997ApJ...489L.179T, 2008ApJ...680.1083R}.  
As noted in Section \ref{DiskICs}, we will use non-cosmological disk simulations to tune up star formation prescription parameters for the different codes, such as the star formation density threshold, the star formation efficiency $\epsilon$, the initial mass of star particles, and the stochasticity of star formation.  

\subsection{Ultraviolet Background}\label{UV}

The metagalactic radiation field provides a lower limit to the intensity of the radiation to which optically thin gas may be exposed. It will be implemented in the {\it AGORA}
simulations using the latest synthesis models of the evolving spectrum of the cosmic UVB by \citet{HaardtMadau12}. Compared to previous calculations \citep{Haardt96,Faucher09}, 
the new models include: {\it 1)} the sawtooth modulation of the background intensity from resonant line absorption in the Lyman series of cosmic hydrogen and helium; {\it 2)} the X-ray emission 
from the obscured and unobscured quasars that gives origin to the X-ray background; {\it 3)} an accurate treatment of the photoionization structure of absorbers that enters in the 
calculation of the helium continuum opacity and recombination emissivity; and {\it 4)} the UV emission from star-forming galaxies following an empirical determination of the star formation 
history of the Universe and detailed stellar population synthesis modeling.  

The resulting UVB intensity has been shown to provide a good fit to the hydrogen-ionization rates inferred from flux decrement and proximity effect measurements, 
predicts that cosmological H~\textsc{ii} (He~\textsc{iii}) regions overlap at redshift 6.7 (2.8), and yields an optical depth to Thomson scattering that is in 
agreement with {\it WMAP} results. If needed, the models will be updated to include, e.g., new measurements of the mean free path of hydrogen-ionizing photons through the IGM
\citep[][but see also \citealt{2013ApJ...765..137O}]{Rudie13}.  

\subsection{Stellar Initial Mass Function and Lifetimes}

In the {\it AGORA} simulations, each star particle represents a simple stellar population with its age, metallicity, and a \citet{Chabrier03} universal IMF, 
\begin{equation}
\phi(m)={dn\over dm}\propto
\begin{cases} e^{-(\log m -\log m_{\rm c})^2/2\sigma^2}/m~~ & (m<1\,\msun)\\
m^{-2.3}~~ & (m>1\,\msun)
\end{cases}
\end{equation}
with $m_{\rm c}=0.08\,\msun$ and $\sigma=0.69$. The IMF is normalized so that $\int m\phi(m)dm=1\,\msun$ between 0.1 and $100\,\msun$.
Star particles will inject mass and metals back into the ISM through Type II and Type Ia SNe explosions, and stellar mass loss.   
{\it The time of this injection depends on stellar lifetimes in the case of Type II and on a distribution of delay times for Type Ia}. 
The former will be determined following the \citet{Hurley00} parameterization for stars of varying masses and metallicities.

\subsection{Metal and Energy Yields of Core-Collapse SNe}\label{Fbck}

We will follow the production of Oxygen and Iron, the yields of which are believed to be metallicity-independent. 
The masses of Oxygen and Iron ejected into the ISM can be converted to a total metal mass as 
\begin{equation}
M_Z = 2.09 M_{\rm O} + 1.06 M_{\rm Fe}, 
\end{equation}
according to the solar abundances of alpha (C, N, O, Ne, Mg, Si, S) and iron (Fe, Ni) group elements of \citet{Asplund09}, as the gas cooling rate is 
a function of total metallicity only. 

\begin{figure}
    \centering
    \includegraphics[width=0.33\textwidth]{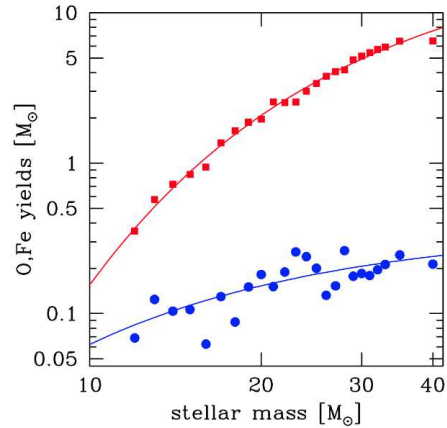}
      \caption{Explosive heavy elements yields of massive stars of solar composition from \citet{Woosley07}. {\it Red squares}: Oxygen. {\it Blue dots}: Iron. 
The solid curves show the best-fitting functions of Eqs.(\ref{eq:MFe}) and (\ref{eq:MO}).
}
\label{fig:yields}
\end{figure}

Stars with masses between 8 and $40\,\msun$ explode as Type II SNe and deposit a net energy of 
$10^{51}\,$ergs into their surroundings. For the assumed IMF, the number of core collapse SNe per unit stellar mass is $0.011\,\msun^{-1}$. 
We will use the following fitting formulae, 
\begin{equation}
M_{\rm Fe}=0.375e^{-17.94/m}~\msun
\label{eq:MFe}
\end{equation}
\begin{equation}
M_{\rm O}=27.66e^{-51.81/m}~\msun
\label{eq:MO}
\end{equation}
to the total mass of ejected Oxygen (including newly synthesized and initial Oxygen) and Iron as a function of stellar mass $m$ (in units of $\msun$) 
tabulated by \citet{Woosley07} for solar metallicity stars (see Figure \ref{fig:yields}). With the assumed IMF, the fractional masses of Oxygen 
and Iron ejected per formed stellar mass are 0.0133 and 0.0011, respectively. For comparison, the mass fractions of Oxygen and Iron in the Sun are 
0.0057 and 0.0013 \citep{Asplund09}.
As discussed in Section \ref{DiskICs}, isolated disk simulations will be used by each code to tune the stellar feedback prescriptions for distributing energy and metals. 

\subsection{Event Rates and Metal Yields of Type Ia SNe}

Type Ia SNe are generally thought to be thermonuclear explosions of accreting carbon-oxygen white dwarfs in close binaries, but the nature of the mass donor star 
remains unknown. To determine how many SNe Ia explode at each timestep, we will adopt the most recent delay time distribution of \citet{Maoz12}. 
The {\it delay times} of SNe Ia are defined as the time intervals between a burst of star formation and the explosion, and follow a power-law $t^{-1}$ in the interval $0.1-10$ Gyr. 
The time integrated number of SNe Ia per formed stellar mass is $0.0013\,\msun^{-1}$, or about 4\% of the stars formed with initial masses in the range, $3-8\,\msun$, 
often considered for the primary stars of SN Ia progenitor systems \citep{Maoz12}. Type Ia SNe leave no remnant, and produce 
\begin{equation}
M_{\rm Fe}=0.63\,\msun~~~~~~~~M_{\rm O}=0.14\,\msun 
\end{equation}
of Iron and Oxygen per event according to the carbon deflagration model W7 of \citet{Iwamoto99}.

\subsection{Mass Loss from Low- and Intermediate-Mass Stars}

Stars below $m=8\,\msun$ return substantial fractions of their mass to the ISM as they evolve and leave behind white dwarf remnants.
In all {\it AGORA} simulations, we will adopt the empirical initial-final mass relation for white dwarfs of \citet{Kalirai08} 
\begin{equation}
w_m=(0.394+0.109m)~\msun 
\label{eq:w_m}
\end{equation}
over the interval $1\,\msun<m<8\,\msun$. We will also assume that stars with $8\,\msun<m<m_{\rm BH}=40\,\msun$ return all but a $w_m=1.4\,\msun$ remnant, and stars above $m_{\rm BH}$ collapse 
to black holes without ejecting material into space, i.e., $w_m=m$.  Few stars form with masses above $40\,\msun$, so the impact of the latter simplifying assumption on chemical evolution is 
minimal. The ``return fraction" -- the integrated mass fraction of each generation of stars that 
is put back into the ISM over a Hubble time -- can be written as 
\begin{equation}
R = \int_{1\,\msun}^{40\,\msun} (m-w_m)\phi(m)dm 
\end{equation}
and is equal to 0.41 for the adopted IMF. Because the return rate is so high, stellar mass losses can prolong star formation even in systems without fresh gas inflow \citep[e.g.,][]{2011ApJ...734...48L, 2011ApJ...738L..24V}.
In practical terms, we shall implement this gas recycling mechanism by determining for each stellar particle the range of stellar masses that die during the current timestep 
and then calculating a returned mass fraction for this mass range using equation (\ref{eq:w_m}). The metallicity of the returned gas is simply the metallicity of the 
star particle, i.e., we will not include metal production by intermediate mass stars. 

\subsection{Notes on AGORA Common Astrophysics}

It is worth briefly noting a few points on the common astrophysics package the {\it AGORA} project has adopted:  
{\it 1)} We are specifying the common astrophysics components because we want these not to be causes for inter-platform differences. Ê
It should be emphasized that we do not aim to determine ``the best'' models to use in galaxy simulations, nor do we attempt to undermine the freedom of choices in the numerical galaxy formation community.  
Any model we adopt here will be outdated soon by better theories and observations.  
We encourage the community to keep developing sophisticated physics and subgrid models for galaxy simulations, and investigate the problem with various methods in an independent manner.  
{\it 2)} Also, the {\it AGORA} common physics package is not about deciding which IMF, energy, or metal yields is the ``correct'' one.  
While our combination of assumptions may allow us to produce ``reasonable, realistic-looking'' galaxies, the sizable number of tunable and/or degenerate parameters makes it less likely to know exactly which assumptions are the ``correct'' ones.  
{\it 3)} As noted in Sections \ref{SFR} and \ref{Fbck}, we do not want to specify at this point the subgrid models for star formation and feedback which we expect to be inevitably different for each code.  
As Section \ref{DiskICs} should make clear, we will use isolated disk initial conditions to tune the models per code, thereby constraining the inter-platform difference.  

\section{Simulations and Analyses}\label{simulations}

The {\it AGORA} galaxy simulation comparison project will proceed via the following steps: {\it 1)} design and perform the multi-platform simulations from common initial conditions and astrophysical assumptions, {\it 2)} examine the simulation outputs on a common analysis platform in a systematic way, and finally {\it 3)} interpret and compare the processed data products across different simulation codes, with strong emphasis on solving long-standing astrophysical problems in galaxy formation.  The guiding strategies for each step of this process are explained in the subsequent sections.  

\subsection{Running Simulations}\label{plans-and-policies}

The {\it AGORA} simulations are designed and planned by the science-oriented working groups in consultation with the {\it AGORA} steering committee, while the simulations themselves will be run by experts on each participating code.   The results of these runs will again be analyzed by members of several science-oriented working groups.  As illustrated in Section \ref{motivation} and \ref{WGs}, the {\it AGORA} project is not a one-time comparison of a set of simulations, but a launchpad to initiate many subprojects, each of which is independently investigated by the science-oriented working groups.  The {\it AGORA} simulations will be run and managed by the following core guidelines.  

\begin{itemize}
\item {\it Designing and running the simulations:}  The {\it AGORA} simulations will be designed with specific astrophysical questions in mind, so that comparing the different simulation outputs can determine whether the adopted astrophysical assumptions are responsible for any success in solving the problem in galaxy formation, rather than artifacts of particular numerical implementations.  The numerical resolution is recommended to be at least as good as that of the {\it Eris} calculation in an attempt to resolve the disk scale height (Sections \ref{need} and \ref{motivation}).  Common initial conditions (Sections \ref{CosICs} and \ref{DiskICs}) and common astrophysical assumptions (Section \ref{physics}) will need to be employed.  Subgrid  prescriptions for stellar feedback will need to be tuned to produce a realistic galaxy in an isolated set-up (Section \ref{DiskICs}). Resolution tests will be encouraged.
\item {\it Data management:}  Each participating code will generate large quantities of unprocessed, intermediate data, in the form of ``checkpoints'' describing the state of the simulation at a given time.  These outputs can be used both to restart the simulation and to conduct analysis.  We plan to store 200 timesteps equally spaced in expansion parameter in addition to redshift snapshots at $z =$ 6, 3, 2, 1, 0.5, 0.2, and 0.0 at the very least.  Each group is also advised to store additional outputs at slightly earlier redshift, shifted by $\Delta z \simeq \pm0.05$.  This extra information may be used to investigate if an inter-platform offset in time-stepping causes ``timing discrepancies'' for halo mergers (see Section \ref{proof-of-concept-results-hmf} for more discussion).  For many timesteps to be analyzed, central data repositories and post-processing compute time will be available at the San Diego Supercomputer Center at the University of California at San Diego, the {\it Hyades} system at the University of California at Santa Cruz, and the {\it Data-Scope} system at the John Hopkins University.  Additionally, we plan to reduce the barrier to entry for the simulation data by making a subset of derived data products available through a web interface.\footnote{The first iteration of {\tt yt} {\it Data-Hub} website is http://hub.yt-project.org/.}
\item {\it Public access:}  One of the key objectives of the {\it AGORA} project is to help interpret the massive and rapidly increasing observational data on galaxy evolution being collected with increasing angular resolution at many different wavelengths by instruments on the ground and in space.  Therefore, a necessary goal of the project is providing access to both direct unprocessed data and derived data products to individuals from the broader astrophysical community.  We intend to make simulation results rapidly available to the entire community,  placing computational outputs on data servers in formats that enable easy comparisons with results from other simulations and with observations.
\end{itemize}

\subsection{Common Analysis}\label{analysis}

Because the simulations are being run in the same dark matter halo merger trees, it is possible to compare the cosmological evolution of target halos predicted by different codes and subgrid assumptions, sometimes halo by halo.  
To this end, the common analysis working group was formed (Section \ref{motivation}) to support the development of common analysis tools, and to define quantitative and physically meaningful comparisons of outputs from all simulation codes.  A key role will be played by the community-developed astrophysical analysis toolkit {\tt yt} \citep{yt, TurkSmith11, Turk13}, which is being instrumented to natively process data from all of the simulation codes being used in the {\it AGORA} project.\footnote{The website is http://yt-project.org/.}

\begin{table*}
\caption{Proof-of-concept Test: Dark Matter-only Simulations of A Galactic Halo of $M_{\rm vir} \simeq 1.7 \times 10^{11}\,\msun$ at $z=0$}
\centering
\begin{tabular}{c || c | c}
\hline\hline 
 & Particle-based Codes & Grid-based Codes \\ 
\hline
Participating codes (Section \ref{proof-of-concept-codes})\tablenotemark{{\it a}} & {\sc Gadget-2/3}, {\sc Gasoline}, {\sc Pkdgrav-2} & {\sc Art-II}, {\sc Enzo}, {\sc Ramses}  \\
\hline
\multirow{2}{*}{Gravitational force resolution (Section \ref{proof-of-concept-setup})} & Force softening of 322 comoving pc until $z=9$, then  &  Finest cell size of 326 comoving pc with \\
 & 322 proper pc from $z=9$ to $z=0$ & adaptive refinement on a particle over-density of 4 \\
\hline
\multirow{3}{*}{Variations in softening (Section \ref{gadget})} & Constant force softening for {\sc Gadget-cfs}  &  \\
& (i.e., 322 comoving pc from $z=100$ to $z=0$), and & Not applicable \\ 
& Adaptive force softening for {\sc Gadget-afs} & \\
\hline
\end{tabular}
\tablenotetext{1}{\scriptsize For detailed explanation, see the referenced sections.}
\label{table:proof-of-concept}
\end{table*}

\begin{itemize}
\item {\it Science-driven analysis:}  {\tt yt} provides a method of describing {\it physical}, rather than {\it computational}, objects inside an astrophysical simulation. For this, {\tt yt} offers tools for selecting regions, applying analysis to regions, visualizing and exporting data to external analysis packages.  {\tt yt} allows astrophysicists to think about the physical questions, rather than the necessary computational steps to ask and answer those questions.
\item {\it Multi-platform analysis:}  In addition to the existing full support for patch-based AMR codes ({\sc Enzo}), {\tt yt} is starting to deliver support for analysis of octree-based AMR outputs ({\sc Art}, {\sc Ramses}) and particle-based outputs ({\sc Gadget}, {\sc Gasoline}). This way, common analysis scripts written in {\tt yt} can be applied to access and investigate data from all of the simulation codes, enabling direct technology transfer between participants, ensuring reproducible scripts and results, and allowing for physically-motivated questions to be asked independent of the simulation platform (see Appendix \ref{app-yt}).
\item {\it Open analysis:}  {\tt yt} is freely available and open source, and is supported by a large community of users and developers (Turk 2013), providing upstream paths for code contribution as well as detailed technical support.  Any newly developed software developed in the project will be naturally available to the broader astrophysical community.  Further, {\tt yt} scripts and the resulting reduced data products can be shared online, enabling data analysis to be conducted by individuals regardless of their affiliation with the project.
\end{itemize}

In our proof-of-concept tests, we have demonstrated that the simulation outputs from all the participating codes of the {\it AGORA} project can be systematically analyzed and visualized in the {\tt yt} platform, using unified, code-independent scripts (see Section \ref{proof-of-concept}).  
Additionally, {\tt yt} can act as input for the {\sc Sunrise} code \citep{Jonsson06, Jonsson+10}, which computes the light from simulated stellar populations (SSPs) and uses ray tracing to calculate the effects of scattering, absorption, and re-emission of this light by dust to generate mock observations and spectral energy distributions (SEDs) of the resulting galaxies in all wavebands.\footnote{The website is http://www.familjenjonsson.org/patrik/sunrise/.}  
The {\sc Sunrise} outputs include realistic images of the simulated galaxies in many wavebands as {\tt FITS} files, which can be compared with observed ones. 
It should be noted that this post-processing inevitably introduces new uncertainties including those in the SSP modeling, or in the calculation of ion number densities. 
Nevertheless, a detailed comparison between high-resolution simulations and the ever-increasing observational data will help constrain the numerical studies of galaxy formation.

\subsection{Issue-based and Science-oriented Comparison of the Simulation Outputs}\label{WGs}

The {\it AGORA} project will consist of a series of {\it issue-based} subprojects, each of which is studied by members of the science-oriented working groups.  As shown in Table \ref{table:science-oriented} of Section \ref{motivation}, these working groups intend to perform original research using multi-platform simulations and produce scientific articles for publication.  Systematic and {\it science-oriented} comparisons of simulation outputs with each other and with observations are highly encouraged, not just a plain code comparison.  For each science question, we will leverage the breadth of the {\it AGORA} simulations -- the varied implementations of subgrid physics, hydrodynamics, and resolution  --  both to understand the differences between simulation outputs and to identify robust predictions. We refer the readers to the {\it AGORA} project website for the scientific objectives of these working groups and the project as a whole.\footnote{See http://sites.google.com/site/santacruzcomparisonproject/details/ or http://www.AGORAsimulations.org/.}

\section{Proof-of-Concept Test}\label{proof-of-concept}

In this section we demonstrate the first, proof-of-concept test of the {\it AGORA} project using a dark matter-only cosmological simulation of a galactic halo of $M_{\rm vir} \simeq 1.7 \times 10^{11}\,\msun$ at $z=0$.  
The primary purpose of this test is to establish and verify the pipeline of the project by ensuring {\it 1)} that each participating code can read in the common ``zoom-in'' initial conditions generated by the {\sc Music} code, {\it 2)} that each code can perform a high-resolution cosmological simulation within a reasonable amount of computing time, and {\it 3)} that the simulation output can be analyzed and visualized in a systematic way using the common analysis {\tt yt} platform.  

\subsection{Experiment Set-up}\label{proof-of-concept-setup}

We design a proof-of-concept dark matter-only simulation with a sub-$L_{\star}$-sized galactic halo described in Section \ref{CosICs}:  a halo of virial mass $M_{\rm vir} \simeq 1.7 \times 10^{11} \msun$ at $z=0$ with a quiescent merger history.  
For a high-resolution ``zoom-in'' simulation of pure dark matter, $\left[ \ell_{\rm min}, \ell_{\rm max} \right] = [7, 12]$ is selected in a $(60\, h^{-1}\, {\rm comoving \,\,Mpc})^3$ cosmological box.  
See Section \ref{CosICs} and Appendix \ref{app-MUSIC} for detailed methods and parameters to generate initial conditions with the {\sc Music} code.
This choice of $\left[ \ell_{\rm min}, \ell_{\rm max} \right]$ corresponds to a dark matter particle resolution of $3.38 \times 10^5 \msun$ in the default highest-resolution region of $2.9 \times 3.9 \times 2.8 \,\,(h^{-1}\, {\rm comoving\,\,Mpc)^3}$.  
In addition, using three variations of the  {\sc Gadget} code ({\sc Gadget-2-cfs}, {\sc Gadget-3-cfs}, and {\sc Gadget-3-afs}; see Section \ref{gadget}), we have tested an initial condition in which the resolution outside the Lagrangian volume of the target halo's $2R_{\rm vir}$ sphere is adaptively lowered (see Section \ref{CosICs} for strategies to minimize the contamination by lower-resolution particles, and to set up a minimum bounding ellipsoid).
The gravitational force softening length of the particle-based codes (e.g., {\sc Gadget-2/3}, {\sc Gasoline}, {\sc Pkdgrav-2}) is set at 322 comoving pc from $z=100$ to $z=9$, and 322 proper pc afterwards until $z=0$ following \cite{2003MNRAS.338...14P}.
Meanwhile the finest cell size of the grid-based codes (e.g., {\sc Art-II}, {\sc Enzo}, {\sc Ramses}) is set at 326 comoving pc, which translates into 11 additional refinement levels in a $2^7$ root grid box (Table \ref{table:proof-of-concept}).   
Cells of the AMR simulations are adaptively refined by factors of 2 in each axis on a dark matter particle over-density of four. 
However we note that the refinement algorithms used in the different AMR codes are not identical, so specifying ``over-density of  four'' does not fully predict the eventual refinements.
Each simulation stores checkpoint outputs at multiple redshifts as described in Section \ref{plans-and-policies} including $z=0$.  

\subsection{Participating Codes}\label{proof-of-concept-codes}

We now briefly explain the participating codes in this test, focusing on the basic architecture of numerical implementations.  
For further information of the groups and participants using each code, we once again refer the interested readers to the {\it AGORA} project website.
The participating codes in the future {\it AGORA} comparison studies are not limited to the ones described herein.

\subsubsection{{\sc Art}}\label{art}

{\sc Art} is an adaptive refinement tree $N$-body+hydrodynamics code that uses a combination of multi-level particle-mesh and shock-capturing Eulerian methods for simulating the evolution of dark matter and gas, respectively.  The code performs refinements locally on individual cells, and cells are organized in refinement trees \citep{Khokhlov1998} designed both to reduce the memory overhead for maintaining a tree and to eliminate most of the neighbor search required for finite-difference operations.  The cell-level, octree-based AMR provides the ability to control the computational mesh on the level of individual cells. Several refinement criteria can be combined with different weights allowing for a flexible refinement strategy that can be tuned to the needs of each particular simulation.  

\begin{itemize}
\item {\sc Art-N}:  {\sc Art} was initially developed as a pure $N$-body code \citep[][]{Kravtsov1997} parallelized for shared memory machines, and later upgraded for distributed memory machines using MPI \citep{Gottloeber2008}.  We denote this code as {\sc Art-N} to differentiate it from the $N$-body + hydrodynamics {\sc Art} code.
\item {\sc Art-I}:  The first shared memory version of the $N$-body + hydrodynamics {\sc Art} was developed in 1998-2001 \citep[][]{Kravtsov1999, Kravtsov2002}. The inviscid fluid dynamics equations are solved using the 2nd-order accurate Godunov method, with piecewise-linear reconstructed boundary states \citep{VanLeer1979}, the exact Riemann solver of \cite{Colella1985}, cooling and heating, and star formation and feedback \citep{Kravtsov2003}.  A version of this code was developed by Anatoly Klypin and collaborators since 2004 with distinct recipes for star formation and feedback \citep[e.g.,][]{Ceverino2009, 2013arXiv1307.0943C}.  These versions will be denoted as {\sc Art-I} in  the {\it AGORA} Collaboration.
\item {\sc Art-II}:  The $N$-body+hydrodynamics {\sc Art} was re-written and parallelized for distributed machines using MPI in 2004 to 2007 \citep{Rudd2008}. It features a flexible time-stepping hierarchy and various physics modules including non-equilibrium H$_2$ formation model \citep{Gnedin2011}, metallicity- and UV-flux-dependent cooling and heating \citep{Gnedin2012}, and sophisticated stellar feedback \citep[e.g.,][]{Agertz2012}.  This code is denoted as {\sc Art-II} in the present and subsequent papers.
\end{itemize}

\subsubsection{{\sc Enzo}}\label{enzo}

{\sc Enzo} is a publicly available AMR code that was originally developed by Greg Bryan and is now driven by community-based development with 32 developers from 14 different institutions over the past four years \citep{Bryan95, BryanNorman1997, OShea2004, 2013arXiv1307.2265T}.\footnote{The website is http://enzo-project.org/.  For the tests described in Section \ref{proof-of-concept-results}, the changeset 99d895b29db1 is used.}   It utilizes the block-structured AMR algorithm of \cite{BergerAMR}.  Dark matter and stars are treated as discrete particles, and their dynamics are solved with the adaptive particle-mesh method \citep{Couchman91}.  To calculate the gravitational potential, Poisson's equation is solved on the root AMR grid with a fast Fourier transform (FFT), and on the AMR grids with a multi-grid relaxation technique.  Here the particle densities are deposited on the AMR grids with a cloud-in-cell interpolation scheme, which is summed with the baryon densities.  The hydrodynamics equations are solved with the 3rd-order accurate piecewise parabolic method \citep{Colella84} that has been modified for hypersonic astrophysical flows \citep{Bryan95}, while multiple Riemann solvers are available to accurately capture shocks within two cells.

\subsubsection{{\sc Gadget-2/3} and Their Variations}\label{gadget}

{\sc Gadget-2} is a three-dimensional $N$-body+SPH code that was developed by Volker Springel as a massively parallel simulation code for distributed memory machines using MPI \citep{Springel01, gadget2}.\footnote{The website is http://www.mpa-garching.mpg.de/gadget/.}  The computational domain is divided between the processors using a space-filling fractal known as a Peano-Hilbert curve to map 3D space onto a 1D curve. This curve can then simply be divided into pieces with each assigned to a different processor. This approach ensures that the force between particles is completely independent of the number of processors, except for the round-off errors. The gravity calculation is performed using a tree method, which organizes the $N$-body particles hierarchically into ``nodes" and approximates the gravitational forces between nodes and particles via a multipole expansion \citep{1986Natur.324..446B}. Time-stepping is done using a kick-drift-kick leapfrog integrator that is fully symplectic in the case of constant timesteps for all particles. To speed up the simulation, individual and adaptive timesteps are employed based on a power-of-two subdivision of the long-range timestep.  

{\sc Gadget-3} is an updated version of {\sc Gadget-2}, and for the purely $N$-body comparison presented here, the two versions are almost equivalent.  However, {\sc Gadget-3}'s improvement in the domain decomposition and dynamic tree reconstruction machinery may induce small but meaningful differences in individual particles' orbits.  {\sc Gadget-3}  was employed in one of the highest resolution ``zoom-in'' collisionless simulations to date, {\it Aquarius} \citep{Aquarius}.  

The proof-of-concept tests also include two variations of gravitational force softening in the {\sc Gadget} code.  {\sc Gadget-cfs} uses the constant gravitational force softening length of 322 comoving pc until $z=0$, different from what other particle-based calculations adopt (Section \ref{proof-of-concept-setup}).  {\sc Gadget-3-afs} employs the same code as {\sc Gadget-3} but with the addition of adaptive force softening lengths in the $N$-body calculation according to the density of the environment, along with a corrective formalism that maintains energy and momentum conservation \citep{Price2007}.  
This has been shown to enhance the clustering of collisionless particles at small scales in cosmological simulations \citep[][we adopt $N_{\rm ngbs} = 90$ for their Eq.(12)]{IannuzziDolag2011}.\footnote{The use of gravitational softening is to limit the spurious two-body interaction noises, since these ``particles'' are in reality interpolation points for smoothed density fields.  
Typically, the softening length is set at a fixed value; however as the system evolves to a highly inhomogeneous structures, the relevance of the choice of softening degrades. 
An algorithm to allow the softening lengths adapt in space and time attempts to circumvent this problem.}  

\begin{figure*}
    \centering
    \includegraphics[width=1.01\textwidth]{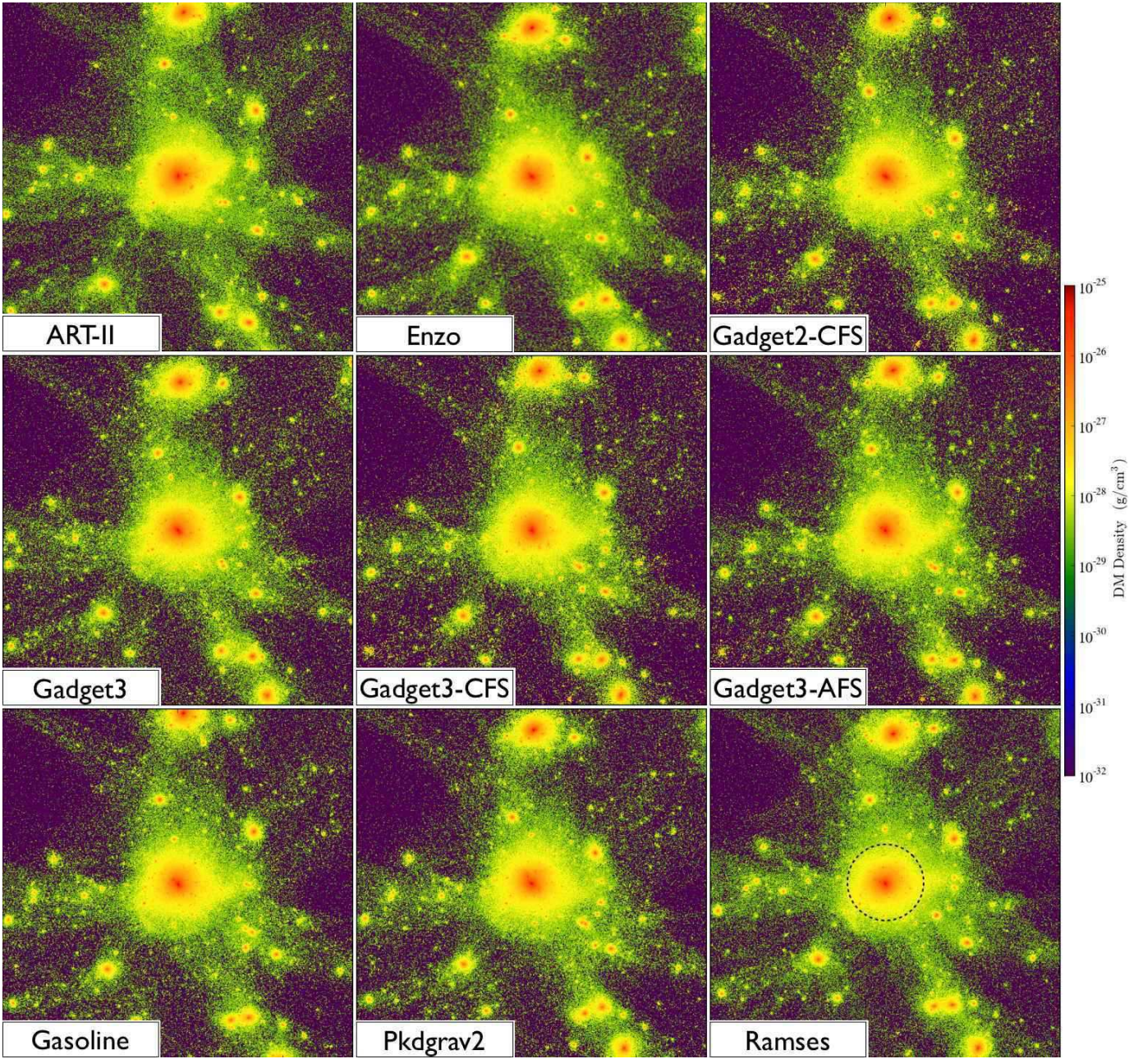}
    \caption{The $z=0$ results of the proof-of-concept dark matter-only tests on a quiescent $M_{\rm vir} \simeq 1.7\times10^{11} \msun$ halo by 9 different versions of the participating codes.  Density-weighted projection of dark matter density in a $1\, h^{-1}\,\,{\rm Mpc}$ box, produced with the common analysis toolkit {\tt yt}. We refer the readers to Table \ref{table:proof-of-concept} and Section \ref{proof-of-concept-codes} for descriptions of the participating codes in this test.  In particular, see Section \ref{gadget} for variations of {\sc Gadget}.  We note that three code groups, -- {\sc Gadget-2-cfs}, {\sc Gadget-3-cfs}, and {\sc Gadget-3-afs} --, have employed an initial condition in which the resolution outside the Lagrangian volume of the target halo's $2 R_{\rm vir}$ sphere is adaptively lowered; see Section \ref{proof-of-concept-setup} for more information.  Hence, particle distributions only within $\sim R_{\rm vir}$ (marked with a {\it dashed circle} in the last panel) can be most reliably compared across all 9 codes with the best available resolution.  Simulations performed by:  Samuel Leitner ({\sc Art-II}), Ji-hoon Kim ({\sc Enzo}), Oliver Hahn ({\sc Gadget-2-cfs}), Keita Todoroki ({\sc Gadget-3}), Alexander Hobbs ({\sc Gadget-3-cfs} and {\sc Gadget-3-afs}), Sijing Shen ({\sc Gasoline}), Michael Kuhlen ({\sc Pkdgrav-2}), and Romain Teyssier ({\sc Ramses}).  The full color version of this figure is available in the electronic edition.  
\label{fig:collage_1Mpc}}
\end{figure*}

\subsubsection{{\sc Gasoline} and {\sc Pkdgrav-2}}\label{gasoline}

{\sc Pkdgrav-2} \citep{stadel01, Ghalo} is a high-performance massively parallel (MPI+pthreads) gravity tree code, employing a fast multipole method \citep[similar to][]{2002JCoPh.179...27D}, but using a 5th-order reduced expansion for faster and more accurate force calculation in parallel, and a multipole-based Ewald summation method for periodic boundary conditions.\footnote{The website is http://hpcforge.org/projects/pkdgrav2/.  For the tests described in Section \ref{proof-of-concept-results}, the changeset e67bd2fd7259 is used.}
Unlike the more commonly employed octree \citep{1986Natur.324..446B}, {\sc Pkdgrav-2} utilizes a binary k-D tree. The tree structure is distributed across processors and is load balanced by domain decomposing the computational volume into spatially local regions, which are adjusted dynamically with each timestep to optimize performance. Particle orbits are calculated with a simple leapfrog integration scheme, using adaptive individual timesteps for particles based on the local dynamical time \citep{2007MNRAS.376..273Z}. The {\sc Pkdgrav-2} code has been used to perform some of the highest resolution collisionless simulations ever performed, {\it Via Lactea II} \citep{VL2} and {\it GHALO} \citep{Ghalo}.

{\sc Gasoline} \citep{gasoline} is a massively parallel $N$-body+SPH code built upon the pure $N$-body  code {\sc Pkdgrav-1}, an earlier version of {\sc Pkdgrav-2}. 
We note that {\sc Pkdgrav-1} uses a timestep criterion that is different from {\sc Pkdgrav-2}'s, one based on the local acceleration rather than the local dynamical time.  
The Ewald summation technique for the long-range force computation is implemented differently in the two codes, too.  
{\sc Gasoline}'s blast-wave and delayed-radiative-cooling stellar feedback \citep[e.g.,][]{stinson06} have been used to produce various types of galaxies, from cored dwarfs \citep{governato10} to Milky Way-like spirals \citep{Eris}.  

\begin{figure*}
    \centering
    \includegraphics[width=0.74\textwidth]{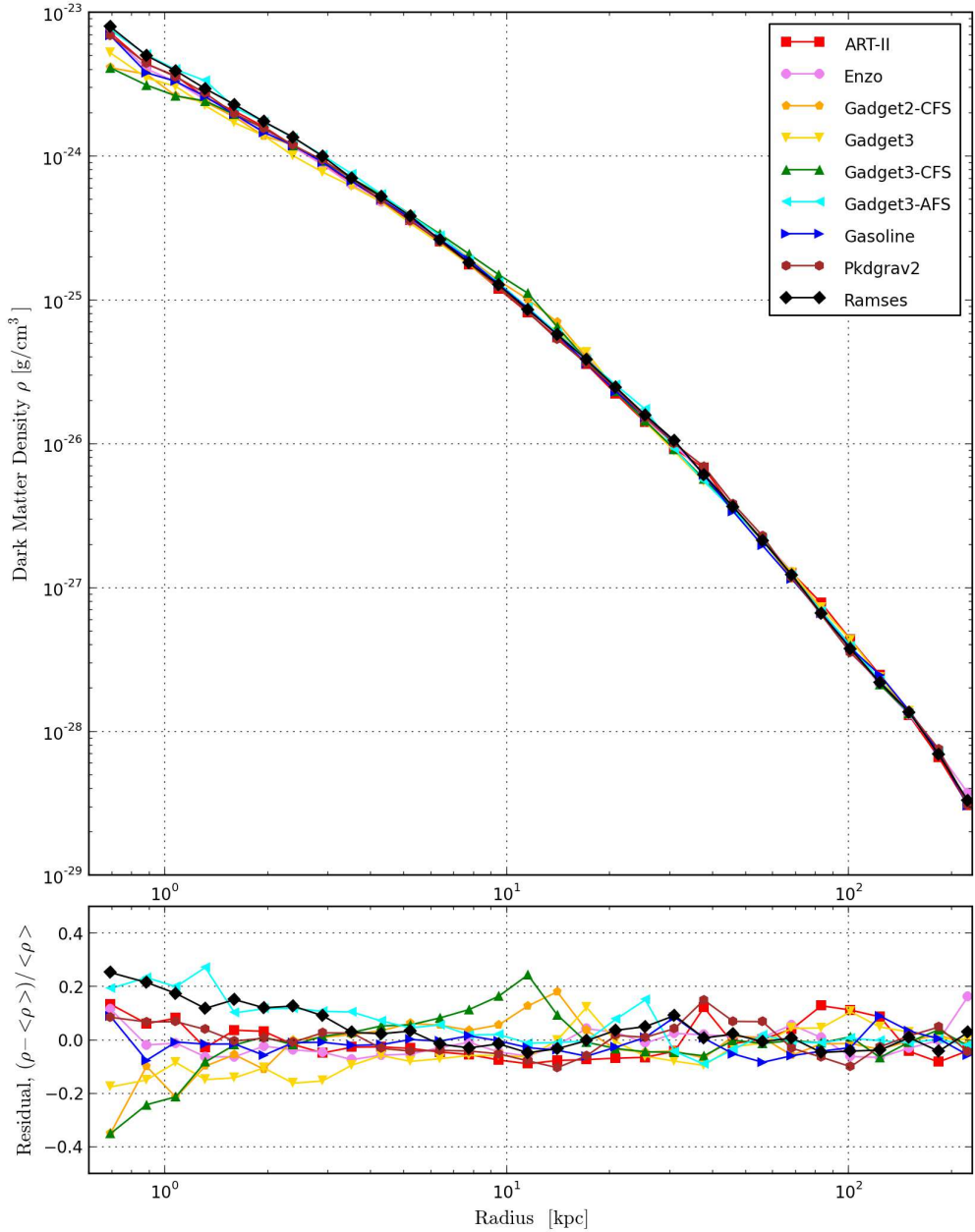}
    \caption{{\it Top:} A composite radial profile of dark matter density centered on the target halo at $z = 0$ formed in the proof-of-concept dark matter-only tests by 9 different versions of the participating codes.  Each profile is generated with the common analysis toolkit {\tt yt}.  {\it Bottom:} fractional deviation from the mean of these profiles.  The full color version of this figure is available in the electronic edition. 
}
\label{fig:composite_profile}
\end{figure*}

\subsubsection{{\sc Ramses}}\label{ramses}

{\sc Ramses} \citep{ramses} is an Eulerian octree-based AMR code that uses the particle-mesh techniques for the $N$-body portion of the calculation and a shock-capturing, unsplit 2nd-order MUSCL scheme (Monotone Upstream-centered Scheme for Conservation Laws) for the fluid component.\footnote{The website is http://www.itp.uzh.ch/$\sim$teyssier/Site/RAMSES.html.}
The Poisson equation is solved on the AMR grid using a multi-grid scheme with Dirichlet boundary conditions on arbitrary domains \citep{2011JCoPh.230.4756G}. 
The fluid can be modeled using the Euler equations, for which various Riemann solvers are implemented (e.g., GLF, HLL, Roe, HLLC and exact). The best compromise between speed and accuracy is offered by the HLLC Riemann solver that we will use in {\it AGORA} simulations \citep{1994ShWav...4...25T}. 
Standard recipes for star formation and stellar feedback are also implemented, the most recent addition being a stellar feedback scheme based on a non-thermal pressure term \citep{2013MNRAS.429.3068T}.

\begin{figure*}
    \centering
    \includegraphics[width=0.96\textwidth]{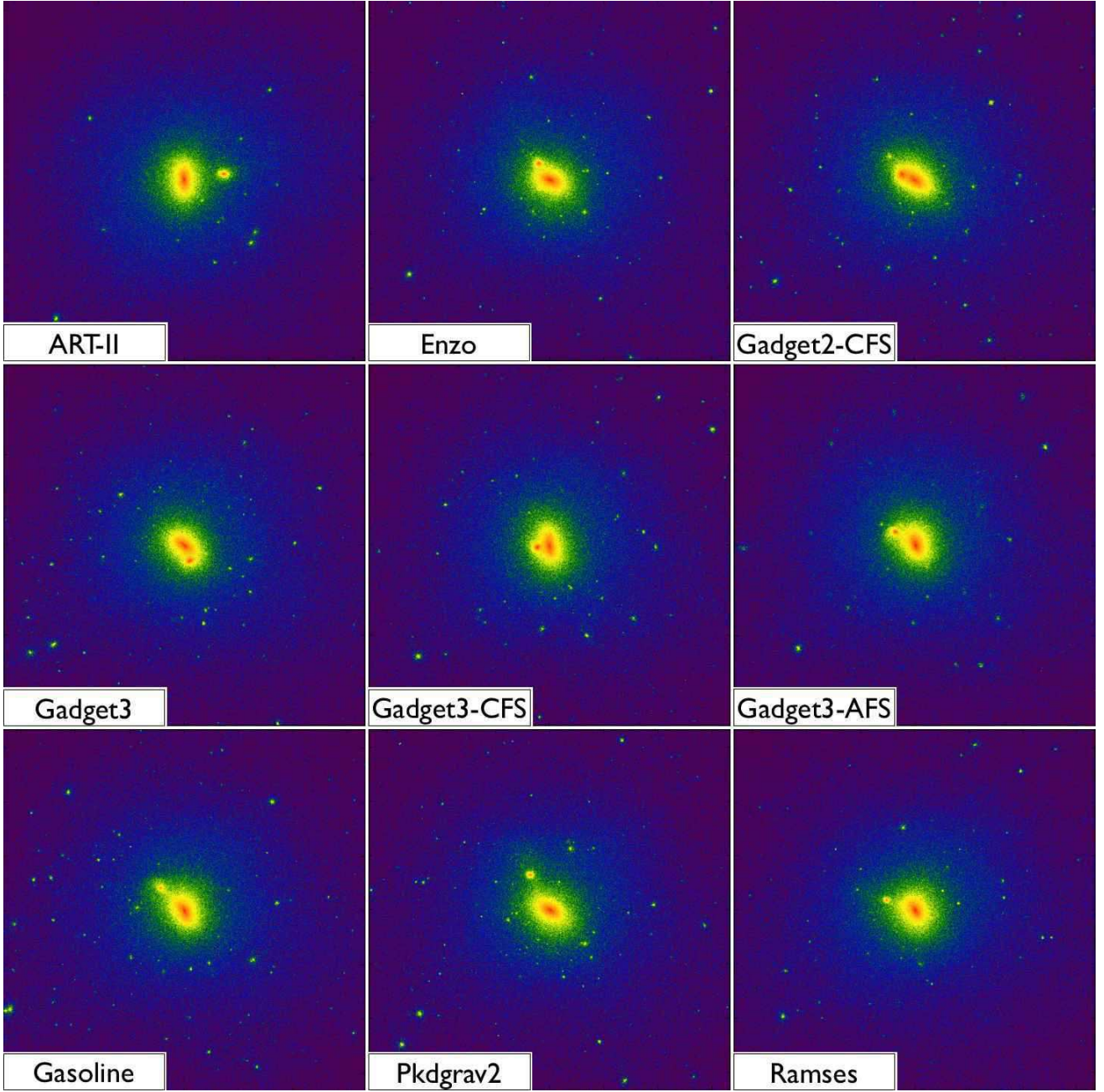}
    \caption{A compilation of 9 maps of density-weighted projection of squared dark matter density from the proof-of-concept dark matter-only tests by 9 different versions of the participating codes  in $200\,\, h^{-1}\,\,{\rm kpc}$ boxes at $z=0$.  The field of view for each panel approximately matches the extent of the virial radius of the host halo ($R_{\rm vir} \simeq 150\,\,{\rm kpc}$).  Panels generated on the common analysis {\tt yt} platform.  For descriptions of the simulation codes and credits, we refer the interested readers to Section \ref{proof-of-concept-codes} and the caption of Figure \ref{fig:collage_1Mpc}.  The full color version of this figure is available in the electronic edition. 
\label{fig:collage_200kpc}}
\end{figure*}

\begin{figure*}
    \centering
    \includegraphics[width=0.71\textwidth]{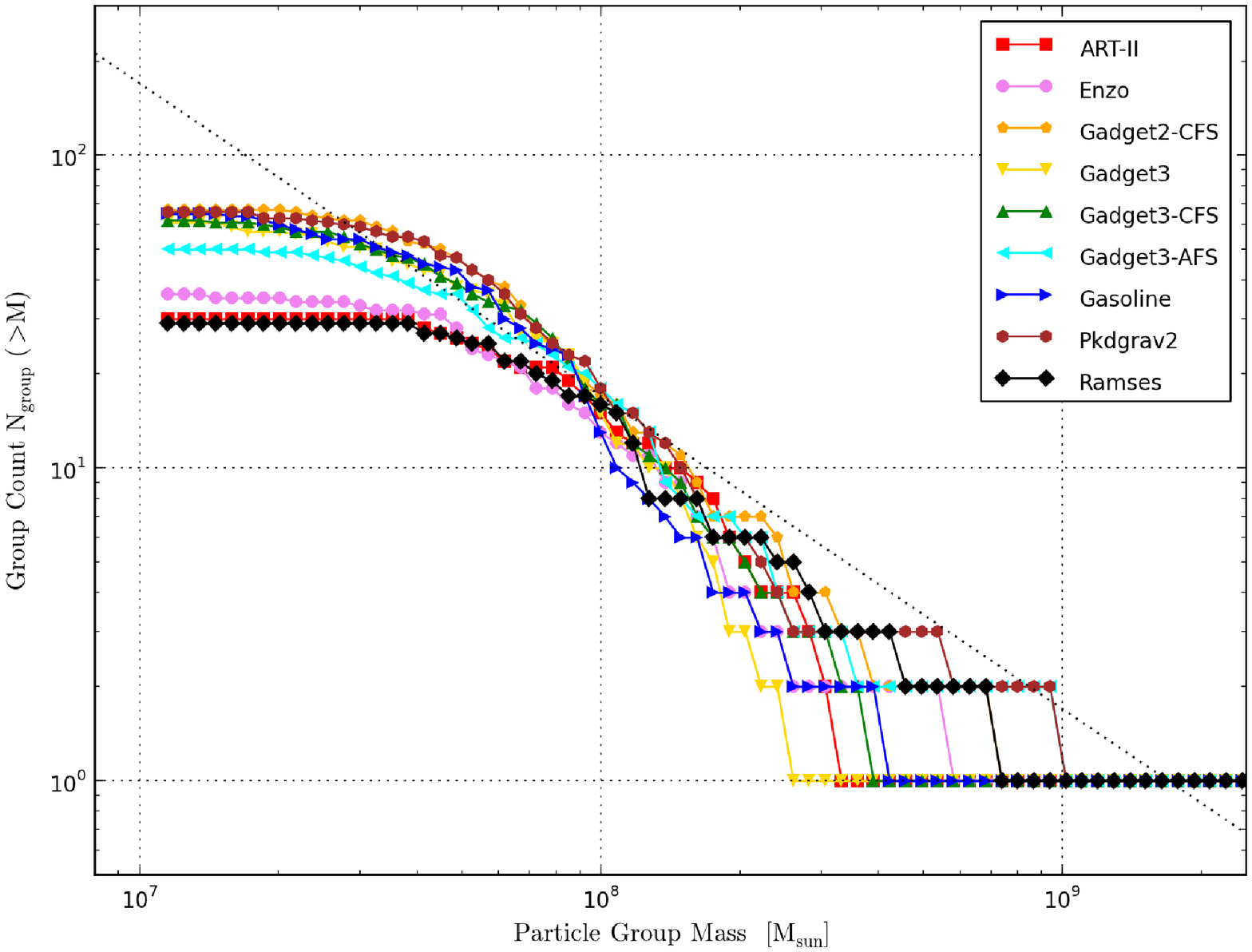}
      \caption{A composite mass function of particle groups identified by the {\sc Hop} halo finder included in {\tt yt} within 150 kpc from the center of the target halo of mass $M_{\rm vir} \simeq 1.7\times10^{11} \msun$ at $z = 0$.  Shown together in a {\it dotted line} is a power-law functional $N(> M) =  0.01(M/M_{\rm host})^{-1}$.   The full color version of this figure is available in the electronic edition. 
}
\label{fig:composite_HMF}
\end{figure*}

\subsection{Results}\label{proof-of-concept-results}

In this section, we lay out the results of the proof-of-concept simulations.  
In particular, the discussion is centered on the basic analysis at the final redshift performed on the common analysis {\tt yt} platform (see Appendix \ref{app-yt} for more information).
More on this dark matter-only simulations including the detailed comparison of the halo catalogues and dark matter merger histories will be presented in the companion paper \citep[][see Section \ref{proof-of-concept-results-discussion}]{IC-paper}.

\subsubsection{Overall Density Structure}\label{proof-of-concept-results-density}

In Figure \ref{fig:collage_1Mpc} we compile 9 image panels that exhibit the results of the proof-of-concept dark matter-only tests by 9 different variations of the participating codes. 
Each panel displays the density-weighted projection of dark matter density in a $1\, h^{-1}\,\,{\rm Mpc}$ box at $z=0$.  
The overall mass distribution around the central halo shows a great similarity across all panels.
The masses of the target halo are also in good agreement with one another, with $\sigma_{\rm M} / \overline M_{\rm vir} \sim 1.2 \%$ from the mean value, $\overline M_{\rm vir}$.  
We, however, caution that three variations of the {\sc Gadget} code ({\sc Gadget-2-cfs}, {\sc Gadget-3-cfs}, and {\sc Gadget-3-afs}; see Section \ref{gadget}) have employed an initial condition in which the resolution outside the Lagrangian volume of the target halo's $2R_{\rm vir}$ sphere is adaptively lowered (see Section \ref{proof-of-concept-setup} for more information).
At this wide field of view, large-scale tidal fields are thus inherently different depending on initial conditions and the aggressiveness of resolution choices in the lower-resolution region.  
Therefore, we remind the readers that particle distributions only within $\sim R_{\rm vir}$ can be most reliably compared across all 9 code platforms with the best available resolution we adopted.

For this reason, we from now on focus only on the structure in the vicinity of the central halo ($R < R_{\rm vir} \,\, \simeq 150 \,\,{\rm kpc}$).  
Assembled in Figure \ref{fig:composite_profile} are dark matter density profiles centered on the target halo of mass $M_{\rm vir} \simeq 1.7\times10^{11} \msun$ at $z=0$.  
To make these profiles, all the dark matter particles within each radial shell are considered, including substructures and lower resolution particles, if any.  
As demonstrated in the bottom panel of Figure \ref{fig:composite_profile}, all 9 profiles agree very well within a fractional difference of 20\% down to a radius of $\sim1$ kpc.  
The location of the maximum density is chosen to be the center of the profile; therefore, the inter-code discrepancies in the centers of profiles may explain the difference in profiles, especially within $\sim 1$ kpc of radius.
We do not find any obvious systematic difference between AMR and SPH codes, or between different gravity solvers.
We again note that all the profiles in this figure are generated with a common {\tt yt} script.
We refer the interested readers to Appendix \ref{app-yt} to see an example script we employed for the presented analysis.  

\subsubsection{Substructure Mass Distribution}\label{proof-of-concept-results-hmf}

Figure \ref{fig:collage_200kpc} shows the density-weighted projections of squared dark matter density of the 9 different proof-of-concept runs at $z=0$ in a $200\,\, h^{-1}\,\,{\rm kpc}$ box.  
It helps to visualize where the substructures are located near the host halo within its virial radius, $R_{\rm vir}$. 
Readers should note that the field of view for each panel approximately encompasses the extent of the virial radius of the target halo ($R_{\rm vir} \simeq 150\,\, {\rm kpc}$ for a $M_{\rm vir} \simeq 1.7\times10^{11} \msun$ halo).  
The structural differences between different code platforms in this scale are more prominent than what is observed in a wider field of view (e.g., Figure \ref{fig:collage_1Mpc}). 
The code-to-code variations at this scale could be attributed to many causes. 
For example, when integrated for a long time, a benignly small deviation in the density distribution at high redshift could evolve into a significant difference later, and become pronounced at $z=0$ especially at this highly zoomed-in scale.  
Because substructures on this scale are in a highly non-linear and dynamically chaotic regime, it would be unlikely to recover halo-to-halo agreements across all platforms.  
A relatively small timing mismatch in the numerical integration of the equations of motion could also prompt a non-negligible disparity when the runs are compared after a long integration.  
Indeed, the discrepancies of the effective timing of the simulations were found to be an important factor in many comparison studies, including the {\it Santa Barbara Cluster Comparison} project \citep[][see also \citealp{gasoline} for further descriptions of the issue in the {\it Santa Barbara} comparison]{SBCluster}.   
These ``timing discrepancy'' precipitates the mismatch in the relative positions of small substructures and in the timing of substructure mergers immediately prior to $z=0$ when we compare the runs.

Another reason for code-to-code variations is the intrinsic difference in numerical methods to solve the Poisson equations for $N$-body dynamics.  
In order to quantitatively inspect such variations, substructures within 150 kpc  from the center (location of the maximum density) of the target halo  are identified by the {\sc Hop} halo finder included in {\tt yt} with an overdensity threshold $\delta_{\rm outer}$ of 80 times the critical density of the Universe \citep{1998ApJ...498..137E, 2010ApJS..191...43S}.\footnote{The websites are http://cmb.as.arizona.edu/$\sim$eisenste/hop/hop.html and http://yt-project.org/doc/analysis$\_$modules/running$\_$halofinder.html.}
The resulting particle group mass functions at $z=0$ are displayed in Figure \ref{fig:composite_HMF}.
Only the groups containing more than 32 particles are drawn.
Shown together in a dotted line is a power-law functional $N(> M) =  0.01(M/M_{\rm host})^{-1}$ that denotes an equal amount of mass per mass decade, to guide the reader's eye.
Note that we have refrained from using the term ``subhalos'' to describe the particle groups identified by {\sc Hop}, because the groups identified this way do not perfectly fit the typical definition of subhalos.  Some of the ``subhalos'' within $R_{\rm vir}$ might have been linked with the host halo by the {\sc Hop} algorithm.

The close resemblances of the mass functions among the particle-based codes with tree-based gravity solvers ({\sc Gadget-2-cfs}, {\sc Gadget-3}, {\sc Gadget-3-cfs}, {\sc Gasoline}, {\sc Pkdgrav-2}) and among the grid-based codes with adaptive meshes ({\sc Art-II}, {\sc Enzo}, {\sc Ramses}) are noticeable.  
However, also unmistakable is the mismatch between these two breeds of codes.  
This phenomenon is indeed well studied and documented by many authors \citep[e.g.,][]{2005ApJS..160....1O, 2005ApJS..160...28H, 2008CS&D....1a5003H}.  
They found that the AMR codes that use a multi-grid or FFT-based gravity solver achieves poorer force resolution at early times than the particle-particle-particle-mesh (${\rm P^3M}$) or tree-PM methods in Lagrangian codes, assuming that the number of base meshes (i.e., grid cells at {\tt levelmax} $\ell_{\rm max} = 12$ in our experiment) is roughly the number of particles, with no or little adaptive mesh at high $z$.  
Due primarily to the lack of force resolution at early redshifts, the low-mass end of the mass function tends to be suppressed for AMR codes.  
Consequently, it has been argued that AMR codes need more resolution in the base grid to achieve the same dark matter mass function at the low-mass end as the Lagrangian codes \citep[e.g.,][]{2005ApJS..160....1O, 2006ApJ...642L..85H}. 
Readers should note, however, the behavior of the adaptive-resolution code {\sc Gadget-3-afs}, which provides results closer to the fixed-resolution codes thanks to its corrective formalism \citep{IannuzziDolag2011}.

We emphasize that the shapes of mass functions may vary not only because of {\it 1)} the intrinsic differences in numerical techniques, but also because of {\it 2)} the inter-platform timing discrepancies discussed earlier, {\it 3)} the force and mass resolution adopted in the test, and {\it 4)} the characteristics of the halo finder.
From these considerations, we argue that it would be premature, if not ill-fated, to characterize a code-to-code difference based solely on the differences in mass functions by a single halo finder at a single epoch.

\subsubsection{Discussion and Future Work}\label{proof-of-concept-results-discussion}

In Section \ref{proof-of-concept} we have presented a conceptual demonstration of the {\it AGORA} project by performing and analyzing a dark matter-only cosmological simulation of a galactic halo of $M_{\rm vir} \simeq 1.7 \times 10^{11}\,\msun$ at $z=0$ with 9 different variations of the participating codes.
We have validated the key infrastructure of the {\it AGORA} project by showing that each participating code reads in the common {\sc Music} initial condition, completes a high-resolution ``zoom-in'' simulation in reasonable time, and provides outputs that can be analyzed in the common analysis {\tt yt} platform.  
Specifically, we point out that all the figures and profiles in Section \ref{proof-of-concept} are generated using unified {\tt yt} scripts that are {\it independent} of the output formats (see, e.g., Appendix \ref{app-yt}).  
Throughout the proof-of-concept test, we have verified the common analysis platform and repeatedly demonstrated its strength.  
For the analyses in future subprojects simple and unified {\tt yt} scripts will be employed, enabling the researchers to focus on physically-motivated questions independent of the simulation codes being analyzed or compared.  

We plan to further investigate these dark matter-only runs in a variety of other dimensions including the comparison of the halo catalogues, dark matter merger histories, and the matter power spectra at various redshifts. 
We also intend to tackle the issue of timing discrepancy so we could obtain the right snapshot that best represents each code for comparison at a given epoch.
We will try to control for this by comparing codes in between their significant mergers, rather than at exactly the same time. Using a merger tree of 5 to 10 most massive substructures as a function of time, we will see whether all codes follow the same sequence of mergers in the same order.  With this information, we will select a redshift for each code - but possibly slightly offset from one another - that is best for inter-platform comparison.
Additionally, in order to correctly quantify the intrinsic code-to-code variations in substructure populations we will study another suite of simulations with higher resolution and see if the discrepancies between mass functions are alleviated. 
Finally, further work and analysis will be performed to identify the halo finder that is the best suitable for future project and for integration within the {\tt yt} platform.
Results from these analyses will be discussed in the forthcoming companion paper \citep{IC-paper}.
 
\section{Summary and Conclusion}\label{conclusion}

Reproducibility is one of the most elementary principles in scientific methods.  
A result from an experiment can be established as scientific knowledge only after the result in its entirety can be reproduced by others within the scientific community according to the same procedure in distinct and independent experimental trials.  
In other words, a conclusion drawn from a single experiment may not be considered as robust until it is verified that the experimental result is not attributed to particular implementations or to an isolated occurrence.
While numerical experiments have become one of the most powerful tools in formulating theories of galaxy formation, it is exactly this requirement of reproducibility that precludes theorists from drawing a definitive conclusion based on a single kind of simulation technique.

Attempts to reproduce the results of numerical experiments in hydrodynamic galaxy simulations, or to compare simulations performed on different platforms, have been hampered by the complexity of the problems including the different assumptions made in different codes regarding the cooling and heating, and subgrid physics and feedback.  
One must strenuously ensure not only that the same physical assumptions are made, but also that identical initial conditions are employed and equivalent quantities are compared across codes.
Because of these reasons, the task of comparing galaxy simulations has been viewed as complex and demanding, even cost-ineffective for researchers.     
The fact that low-resolution ($>$ kpc) galaxy simulations inevitably introduce phenomenological recipes to describe stellar subgrid physics that are heavily dependent on code characteristics only compounds the problem.

The {\it AGORA} project is a collective response of the numerical galaxy formation community to such a challenge.  
It is an initiative to promote a {\it multi-platform} approach to the problems in galaxy formation from the beginning, which is essential to verify that astrophysical assumptions are accountable for any success, not particular simulation techniques or implementations.  
To this end, in this paper we have developed the framework of the project, and introduced its principal components.
First, we have created the common cosmological and isolated initial conditions for the {\it AGORA} simulations (Section \ref{ICs}).  
Two sets of cosmological halos are identified using the {\sc Music} code, one with quiescent and the other with violent merger histories.  
A set of isolated disk initial conditions of varying mass resolution is also built  with which subgrid stellar physics will be tuned for each code to produce realistic galaxies.
We have also established the common astrophysical assumptions to be utilized in all of {\it AGORA} hydrodynamic simulations based on the consensus among the codes participating in the comparison (Section \ref{physics}).  
The package includes the metallicity-dependent gas cooling, UV background, stellar IMF, star formation, metal and energy yields by supernovae, and stellar mass loss.  
Lastly, we have constructed the common analysis platform on the open source {\tt yt} code, which will play an imperative role in the project as it takes as input the outputs from all of the participating codes (Section \ref{analysis}).
Building of the {\it AGORA} infrastructure has been driven by the task-oriented working groups  whose goal is to ensure that the {\it AGORA} comparisons are meticulously bookended by common initial conditions, common astrophysics, and common analysis (Table \ref{table:task-oriented}). 

In order for the {\it AGORA} project to be maximally useful in addressing the outstanding problems in galaxy formation, we argue that achieving high, state-of-the-art numerical resolution is important as the interplay between resolution and subgrid prescriptions is a key component in modeling galaxy formation (Section \ref{need}).   
This way we aim to better understand and lift the degeneracies between subgrid treatments of contemporary galaxy formation simulations.
The simulation data at multiple epochs will be stored for analysis and reproducibility, and will be publicly available to the community for fast access (Section \ref{plans-and-policies}).
We also present the {\it AGORA} project as a stage platform for further galaxy formation studies by encouraging science-oriented and issue-based comparisons of simulations using the infrastructure developed here (Section \ref{WGs}).  
Indeed, the project already serves as a launchpad to initiate many science-oriented subprojects in the {\it AGORA} Collaboration (Table \ref{table:science-oriented}).  

To field-test the {\it AGORA} infrastructure, proof-of-concept dark matter-only simulations of a galactic halo with a $z=0$ mass of $M_{\rm vir} \simeq 1.7 \times 10^{11}\,\msun$ have been conducted by 9 variations of the participating codes (Section \ref{proof-of-concept}).
We have found that the dark matter density profiles as well as the general distributions of matter exhibit good agreement across codes, providing a solid foundation for future hydrodynamic simulations.  
Throughout the test we have demonstrated the practical advantage of our common initial conditions and analysis pipeline by showing that each code can read the identical ``zoom-in'' {\sc Music} initial conditions (e.g., Appendix \ref{app-MUSIC}) and that each simulation output can be analyzed with a single {\tt yt} script independent of the output format (e.g., Appendix \ref{app-yt}).  
By doing so, we have produced evidence that the cumbersome barriers in comparing galaxy  simulations can be, and are, removed.  
The framework we assembled for the {\it AGORA} project will allow the numerical galaxy formation community to routinely and expeditiously compare their results across code platforms, collectively raising the  predictive power of numerical experiments in galaxy formation.  

As the discussion in Section \ref{WGs} should make clear, this paper will be followed by many science-oriented studies of galaxy simulations that leverage the breadth of participating codes in the {\it AGORA} project.
We will tackle long-standing challenges of cosmological galaxy formation by systematically comparing simulations using different codes and different subgrid prescriptions with each other and with observations.   
We also emphasize that the {\it AGORA} project is an open platform, and we encourage any interested individuals or groups to participate.
For instance, the scope of simulation codes that will partake in future {\it AGORA} comparisons is not limited to those that are described in this paper.
Notably, different flavors of SPH such as {\sc Gadget-3-sphs} \citep{2012MNRAS.422.3037R} will be included in {\it AGORA} hydrodynamic simulations.   
Code groups such as {\sc Art-I} (Section \ref{art}) and {\sc Nyx} \citep{AlmBell12}  have already verified  that they can import the common initial conditions of the project, and analyze their outputs in the common analysis {\tt yt} platform.

\vspace{2mm}

The authors of this article thank members of the {\it AGORA} Collaboration who are not on the author list but have provided helpful suggestions on the early version of the paper, including Peter Behroozi, Romeel Dav\'{e}, Michele Fumagalli, Fabio Governato, and Ramin Skibba. 
We thank Volker Springel for private communication on the contents of Section \ref{gadget} and providing the original version of {\sc Gadget-3}. 
We thank Joachim Stadel and Doug Potter for private communication on the contents of Section \ref{gasoline} and providing the original version of {\sc Pkdgrav-2}.
We gratefully acknowledge the financial and logistical support from the University of California High-Performance AstroComputing Center (UC-HiPACC) during the two {\it AGORA} workshops held at the University of California Santa Cruz in 2012 and 2013.  
Ji-hoon Kim and Mark R. Krumholz acknowledge support from NSF through grant AST-0955300, NASA through grant NNX13AB84G, and a Chandra Space Telescope grant GO2-13162A.
Ji-hoon Kim thanks the additional support from the UC-HiPACC.  
He is also is grateful for the support from Stuart Marshall and the computational team at SLAC National Accelerator Laboratory during the usage of the {\it Orange} cluster for the generation and testing of the {\it AGORA} initial conditions, and for the support from Shawfeng Dong and the computational team at the University of California Santa Cruz during the usage of the {\it Hyades} cluster for the analysis of the {\it AGORA} proof-of-concept runs.
Avishai Dekel and Adi Zolotov acknowledge support by ISF grant 24/12, by GIF grant G-1052-104.7/2009, by a DIP grant, by NSF grant AST-1010033, and by the I-CORE Program of the PBC and the ISF grant 1829/12.
Nathan J. Goldbaum acknowledges support from NSF grant AST-0955300 and the Graduate Research Fellowship Program.
Oliver Hahn acknowledges support from the Swiss National Science Foundation through the Ambizione Fellowship.
Samuel N. Leitner acknowledges support by an Astronomy Center for Theory and Computation Prize Fellowship at the University of Maryland.
Piero Madau acknowledges support from NSF through grants OIA-1124453 and AST-1229745.
Kentaro Nagamine and Keita Todoroki's computing time was provided by XSEDE allocation TG-AST070038N and they utilized the Texas Advanced Computing Center's {\it Lonestar}.  
XSEDE is supported by NSF grant OCI-1053575.
Jose O\~{n}orbe acknowledges the financial support from the Fulbright/MICINN Program and NASA grant NNX09AG01G. 
His computing time was provided by XSEDE allocation TG-AST110035.
Brian W. O'Shea and Britton D. Smith acknowledge support of the LANL Institute for Geophysics and Planetary Physics, NASA through grants NNX09AD80G and NNX12AC98G, and by NSF through grants AST-0908819, PHY-0941373, and PHY-0822648.  
Their computing time was provided by XSEDE allocations TG-AST090040 and TG-AST120009.  
Brian W. O'Shea's work was supported in part by NSF through grant PHYS-1066293 and the hospitality of the Aspen Center for Physics.
Joel R. Primack acknowledges support from NSF grant AST-1010033.
Thomas Quinn acknowledges support from NSF grant AST-0908499.
Justin I. Read acknowledges support from SNF grant PP00P2\_128540/1.
Douglas H. Rudd acknowledges support from NSF grant OCI-0904484, the Research Computing Center and the Kavli Institute for Cosmological Physics at the University of Chicago through NSF grant PHY-1125897 and an endowment from the Kavli Foundation and its founder Fred Kavli.  
His work made use of computing facilities provided by the Research Computing Center at the University of Chicago, the Yale University Faculty of Arts and Sciences High Performance Computing Center, and the Joint Fermilab - KICP Supercomputing Cluster, supported by grants from Fermilab, Kavli Institute for Cosmological Physics, and the University of Chicago.
Romain Teyssier and Oliver Hahn's {\sc Ramses} simulations were performed on the Cray XE6 cluster {\it Monte Rosa} at CSCS, Lugano, Switzerland.
Matthew J. Turk acknowledges support by the NSF CI TraCS Fellowship award OCI-1048505.
John H. Wise acknowledges support from NSF grant AST-1211626.

\begin{appendix}

\section{A. Common Initial Condition Generator Music: Example Parameter for Cosmological Runs}\label{app-MUSIC}

The following {\sc Music} parameter file produces a cosmological initial condition that is used in the proof-of-concept dark matter-only test described in Section \ref{proof-of-concept}. 
By simply modifying the {\tt [output]} parameters one can generate initial conditions for various other simulation codes.

\vspace{2mm}

{\footnotesize

${\tt [setup]}$

${\tt boxlength	= 60}$

${\tt zstart	= 100}$

${\tt levelmin	= 7}$

${\tt levelmin\_TF = 9}$

${\tt levelmax	= 12}$

${\tt padding		= 9}$

${\tt overlap		= 4}$

${\tt ref\_offset		= 0.618,\,\, 0.550,\,\, 0.408}$

${\tt ref\_extent		= 0.048,\,\, 0.065,\,\, 0.047}$

${\tt align\_top		= yes}$

${\tt periodic\_TF	= no}$

${\tt baryons		= no}$

${\tt use\_2LPT		= no}$

${\tt use\_LLA		= no}$

${\tt center\_vel	= no}$ \newline

${\tt [cosmology]}$

${\tt Omega\_m	= 0.272}$

${\tt Omega\_L		= 0.728}$

${\tt Omega\_b		= 0.0455}$

${\tt H0			= 70.2}$

${\tt sigma\_8		= 0.807}$

${\tt nspec		= 0.961}$

${\tt transfer		= eisenstein}$ \newline

${\tt [random]}$

${\tt cubesize		= 256}$

${\tt seed[8]		= 95064}$

${\tt seed[9]		= 31415}$

${\tt seed[10]		= 27183}$ \newline

${\tt [output]}$

${\tt format		= enzo}$

${\tt filename		= ic.enzo}$ \newline

${\tt [poisson]}$

${\tt fft\_fine		= yes}$

${\tt accuracy		= 1e-4}$

${\tt grad\_order	= 6}$

${\tt laplace\_order	= 6}$

}
\vspace{2mm}

For more information on the common cosmological initial conditions of the {\it AGORA} project and its primary tool {\sc Music} \citep{MUSIC}, see Section \ref{CosICs} and the {\sc Music} website http://bitbucket.org/ohahn/music/.  

\section{B. Common Analysis Platform yt: Example Script}\label{app-yt}

The following {\tt yt} script written in {\tt python} generates a radial profile of enclosed dark matter mass   from which plots like Figure \ref{fig:composite_profile} can be derived.  
This script works for various simulations outputs including all represented in the proof-of-concept study (Section \ref{proof-of-concept}) with the development tree of {\tt yt-3.0}. 

\vspace{2mm}

{\footnotesize

${\tt from \,\,\,\, yt.mods \,\,\,\,  import \,\,\,\,  *}$

${\tt import \,\,\,\,matplotlib.pyplot\,\,\,\,  as \,\,\,\, plt}$

${\tt ds = load(''DD0040/data0040'')}$

${\tt radius1 = 0.8}$

${\tt radius2 = 300}$

${\tt total\_bins = 30}$ \newline

${\tt sp = ds.h.sphere([0.5, 0.5, 0.5], \,\,(radius2,\,\, 'kpc'))}$

${\tt prof = BinnedProfile1D(sp, \,\,total\_bins, \,\,''ParticleRadiuskpc'', }$

${\tt \ \ \ \ \ \ \ \ \ \ \ \ \ \ \ \ \ \ \ \ \ \  radius1, \,\,radius2, \,\,end\_collect = True)}$ 

${\tt prof.add\_fields([(''all'', \,\,''ParticleMassMsun'')], \,\,weight = None, }$

${\tt \ \ \ \ \ \ \ \ \ \ \ \ \ \ \ \ \ \ \ \ \ \  accumulation=True)}$ 

${\tt plt.loglog(prof[''ParticleRadiuskpc''], }$ 

${\tt \ \ \ \ \ \ \ \ \ \ \ \ \ \ \ \ \ \ \ \ \ \  prof[(''all'', \,\,''ParticleMassMsun'')], \,\,'-k')}$

${\tt plt.xlabel(''Radius\,\,\,\, [kpc]'')}$

${\tt plt.ylabel(''Enclosed \,\,\,\, Dark\,\,\,\, Matter\,\,\,\, Mass\,\,\,\, [Msun]'')}$

${\tt plt.savefig(''\%s\_encmass.png'' \,\,\% \,\,ds)}$

}
\vspace{2mm}

Interested readers may want to try an extended version of the unified {\tt yt} script at http://bitbucket.org/mornkr/agora-analysis-script/ employed in the analyses of the proof-of-concept runs. 
For the analysis described in Section \ref{proof-of-concept-results}, the {\tt yt-3.0} changeset e018996fcb31 is used.  
For more information on the common analysis philosophy of the {\it AGORA} project and its toolkit {\tt yt}  \citep{yt}, see Section \ref{analysis} and the {\tt yt} website http://yt-project.org/.

\end{appendix}

\end{document}